\documentstyle[11pt,twoside,fleqn,psfig]{article}
\newcounter{mycounter}
\newenvironment{my_item}{\begin{list}{$\bullet$}{\setlength{\itemsep}{0.1mm}
\setlength{\parsep}{0.1mm} \setlength{\topsep}{0.1mm}
\setlength{\rightmargin}{\leftmargin}}}{\end{list}}

\font\BbbNormal = msbm10
\font\BbbScript = msbm7
\newfam\Bbbfam 
\textfont\Bbbfam=\BbbNormal
\scriptfont\Bbbfam=\BbbScript

\begin{document}
\begin{center}
{\huge \bf Small-worlds: How and why}
\end{center}

\begin{center}
{\large  Nisha Mathias$^{1}$ and Venkatesh Gopal$^{2}$}
\end{center}

\begin{center}
$^{1}${\it Department of Computer Science and Automation, 
Indian Institute of Science,\\ Bangalore 560012, INDIA; email:
nisha@csa.iisc.ernet.in}
\end{center}
\vspace{-1cm}
\begin{center}
$^{2}${\it Raman Research Institute, Sadashivanagar, Bangalore
560080, INDIA; email: vgopal@rri.ernet.in}
\end{center}
\date{}

\begin{abstract}
We investigate small-world networks from the point of view of their origin. 
While the characteristics of small-world networks are now fairly well 
understood, there is as yet no work on what drives the emergence of such a
network architecture. In situations such as neural or transportation 
networks, where a physical distance between the nodes of the network exists,
we study whether the small-world topology arises as a consequence of a 
tradeoff between maximal connectivity and minimal wiring. Using simulated
annealing, we study the properties of a randomly rewired network as the
relative tradeoff between wiring and connectivity is varied. When the
network seeks to minimize wiring, a regular graph results. At the other
extreme, when connectivity is maximized, a near random network is obtained. 
In the intermediate regime, a small-world network is formed. However, 
unlike the model of Watts and Strogatz (Nature {\bf 393}, 440 (1998)), 
we find an alternate route to small-world behaviour through the formation
of hubs, small clusters where one vertex is connected to a large number
of neighbours. 

\end{abstract}
\section{Introduction}
Coupled systems may be modelled as networks or graphs, where the
vertices represent the elements of the system, and the edges represent
the interactions between them. The topology of these networks
influences their dynamics. Network topologies may be random, where
each node or vertex is randomly wired to any other node; or they may
be regular, with each vertex being connected to a fixed number of
neighbouring nodes. Watts and Strogatz \cite{3:watts1} showed that
between these two extremes lay another regime of connectivity, which
they called a {\em small-world} network. Such networks are `almost'
regular graphs, but with a few long range connections.

What does it mean to have a `long-range' connection?  Consider a few
examples of networks: neurons in the brain, transportation and social
networks, citations of scientific papers and the world wide web.
There is a difference between the elements of this list. Social networks, 
paper citations and the internet, are networks where the
links have no physical distance. For example, a link between two
websites physically far apart is no different from one between two
machines that are next to one another. Neural and transportation networks
however, have a well defined physical distance between their nodes.
In this paper, we investigate how placing a {\em cost} on the length of an
edge affects the connectivity of the network.

We now briefly describe the small-world model of Watts and Strogatz
(WS) and also introduce the notation that we shall
use. WS considered a ring lattice; $n$ sites arranged at regular
intervals on a ring, with each vertex connected to
$k$ nearest neighbours. Disorder
is introduced into the graph by randomly rewiring 
each of the edges with a probability $p$. While at $p = 0$, the graph 
remains $k$-regular, at $p = 1$, a random graph results.
They quantified the structural properties of this lattice by two 
parameters, $L$ and $C$. $L$, the {\em characteristic path length} 
reflects the average connectivity of the network, while $C$, the 
{\em clustering coefficient}  measures the extent to which neighbours
of a vertex are neighbours of each other. Networks exhibiting
small-world behaviour are characterized by low characteristic path length, and
high clustering coefficient. Finally, we point out
that there are two kinds of distances in a graph. One is the {\em graph}
distance, the minimal number of links between any two vertices
of the graph. The other is the Euclidean or {\em physical} distance
between these vertices.

Although recent work has shown small-worlds to be pervasive in a 
range of networks that arise from both natural and man-made technology 
\cite{3:watts1,3:barabasi}, the hows and whys of this ubiquity 
have not been explained. The fact
that small-worlds seem to be one of nature's `architectural'
principles, leads us to ask what constraints might force networks to
choose a small-world topology. We attempt to understand the emergence
of the small-world topology in networks where the physical distance is
a criterion that cannot be ignored.

\section{Can small-worlds arise as the result of an Optimization?}

Consider a toy model of the brain. Let us assume that it consists of
local processing units, connected by wires. What constraints act on
this system?  On the one hand, one would want the highest connectivity
between the local processing units so that information could be
exchanged as fast as possible. On the other, it is wasteful to wire
everything to everything else. The energy requirements are higher,
more heat is generated, and more material needs to be used, and
consequently, more space is occupied. Unrealistic though this model
is, it motivated us to examine whether small-worlds would emerge as
the result of these constraints.

The concept of multiple scales was introduced by Kasturirangan
\cite{3:kasturi}, where he asserted that the fundamental mechanism
behind the small-world phenomena is not disorder or randomness, but
the presence of edges of many different length scales. The {\em length
scale} of a newly introduced edge $e_{ij}$, is defined to be the graph
distance between vertices $i$ and $j$ {\em before} the
edge was introduced. He argued that the distribution of length scales
of the new edges is significantly more important than whether the new
edges are long, medium or short range.

\begin{figure}[!htbp]
\centerline{\psfig{figure=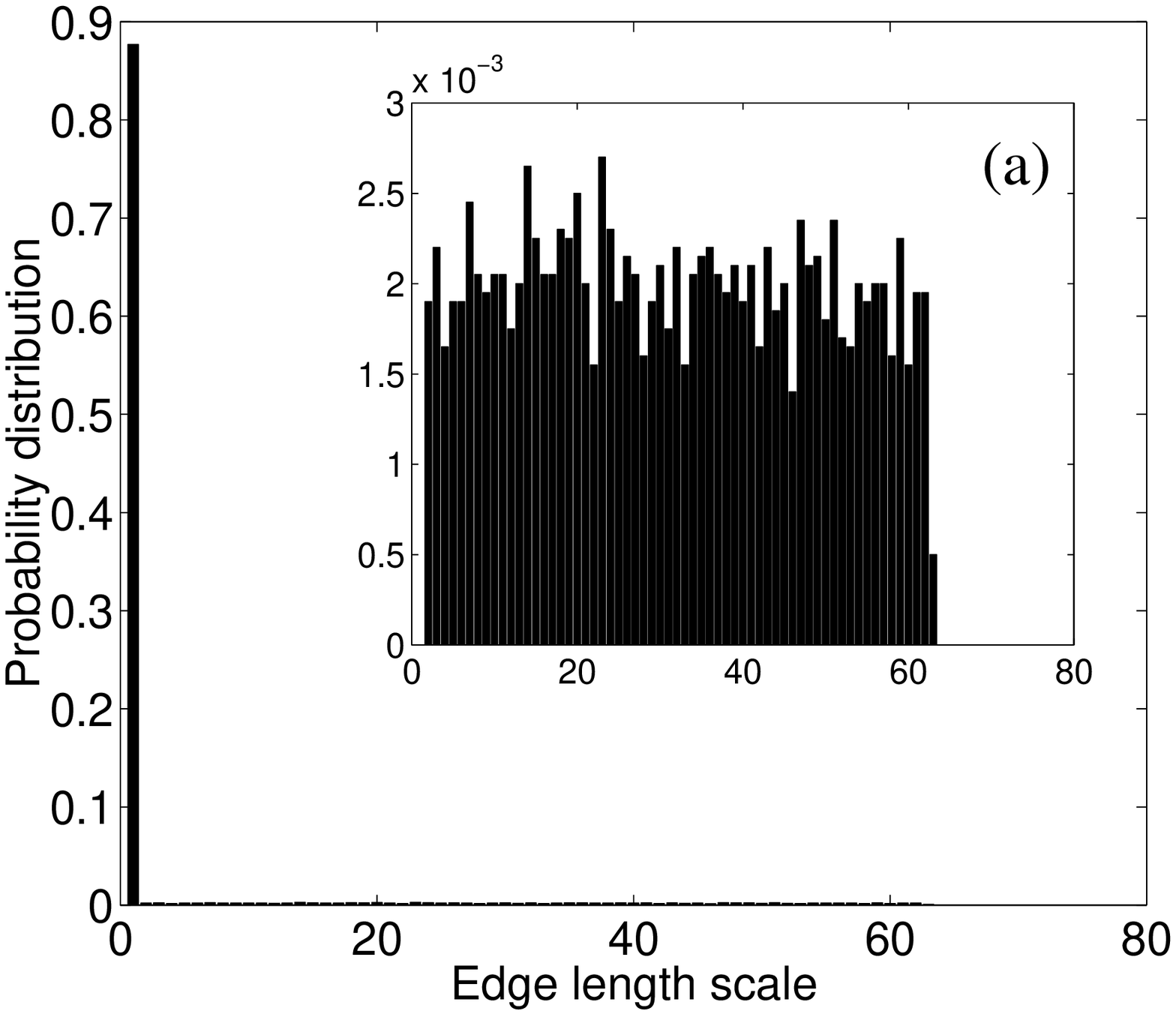,width=7.3cm,height=7.3cm} 
            \psfig{figure=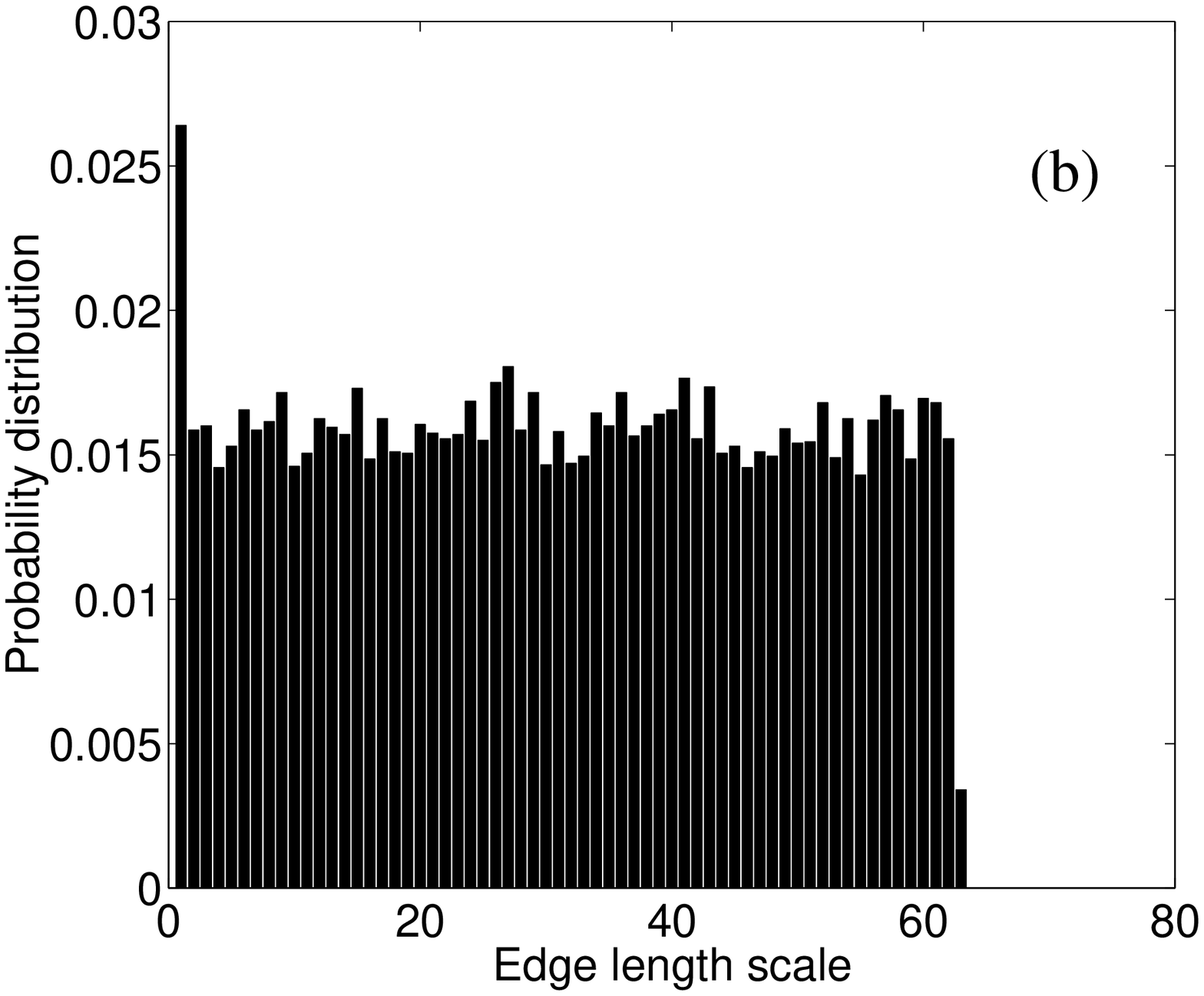,width=7.3cm,height=7.3cm}}
\caption{Edge length scale distribution at (a) $p$=0.125, and (b) $p$=1.00; 
$p$ being the degree of disorder introduced into the $n$=250, $k$=4 
regular network using the WS rewiring procedure to introduce small-world
behaviour. The inset in
(a) displays the distribution of all length scales other than the unit scale.
Both plots are averaged over 25 samples.} 
\label{fig:small-world and random network}
\end{figure}

We obtain the edge scale distribution by binning the length scales of all the
edges in a graph, with respect to its corresponding regular graph.
Starting with a $k$-regular graph, and using the WS
rewiring procedure,  
we study the edge scale distribution at various degrees of disorder. 
Figure~\ref{fig:small-world and random network} shows the edge scale
distribution at two degrees of disorder. Figure~\ref{fig:small-world and 
random network}(a) shows the edge
scale distribution in the small-world regime, (p = 0.125). Due to
introduction of a small amount of disorder, a few edges are rewired to
become far and consequently have a large length scale. However,
they are too few in number to significantly alter the length scale
distribution and hence, the edges of unit length scale dominate the
distribution. Figure~\ref{fig:small-world and 
random network}(b) shows the edge scale distribution at $p=1$,
a random graph. Here, the edges are uniformly distributed over the
entire length scale range, that is, from 1 to $n/k$. The network
still retains a slight bias towards the unit length scale. At both
these degrees of randomness however, the characteristic path length
scales logarithmically with $n$. There thus appears to be some factor
that constrains the distribution of edge length scales to (a) and not
(b), namely, restricting the rewiring to just a few far edges. We
question whether the association of a cost to each edge, proportional
to its length, serves to work as this constraint.

\section{Optimization model}

We use the method of simulated annealing
\cite{3:press} to find the network which results
in the best optimization of the objective function $E$, whose
minimization is the goal of the procedure. The network used in the
model is that of vertices arranged symmetrically along a ring. The
size of the network, $n$, as well as the total number of edges are
fixed. So also are the positions of the vertices, which are equally
spaced along the circumference of the circle. Initially, the network
is $k$-regular, similar to the WS model. The configuration has an
associated energy $E$, a function of both its wiring cost and the
average degree of separation between its vertices. The
objective function $E$ is taken to be, 
\begin{eqnarray*} E &= &\lambda L + (1-\lambda)W, \end{eqnarray*} 
a linear combination of the
normalized characteristic path length $L$, and the normalized wiring
cost $W$. The characteristic path length $L$, as defined by Watts and
Strogatz, is the average distance between all pairs of vertices, given
by \begin{eqnarray*} L &= &\frac{1}{n(n-1)} \sum \limits_{i \neq j}
d_{ij} , \end{eqnarray*} where $d_{ij}$ is the number of links along
the shortest path between vertices $i$ and $j$. It is therefore a
measure based on  graph distance, and reflects the global
connectivity among all vertices in the graph. The wiring cost $W$, in
contrast, is a measure of the physical distance between
connected vertices. The cost of wiring an edge $e_{ij}$, is taken to
be the Euclidean distance between the vertices $i$ and $j$. Hence,
the total wiring cost is 
\begin{eqnarray*} W &= &\sum \limits_{e_{ij}}
\sqrt{(x_i - x_j)^2 + (y_i - y_j)^2 }, \end{eqnarray*} 
where
$(x_i,y_i)$ are the coordinates of vertex $i$ on the ring lattice.
The characteristic path length $L$ is normalized by $L(0)$, the path
length in the $k$-regular network; while $W$ is normalized by the
total wiring cost that results when the edges at each vertex are the
longest possible, namely, when each vertex is connected to its
diametrically opposite vertex, and to the vertices surrounding it.
The parameter $\lambda$ is varied depending on the relative importance
of the minimization of $L$ and $W$. One can regard $(1-\lambda)$
as the wiring cost per unit length, and $W$ as the length of
wiring required.

Starting from the initial regular network, a standard Monte Carlo
scheme \cite{3:press} is used to search for the energy minimum.
Similar to the WS model, duplicate edges and loops were not allowed,
and it was ensured that the rewiring did not result in isolated
vertices. The starting value for $T$, the annealing `temperature',
was initially chosen to be the initial energy, $E$, itself. The 
temperature was then lowered in steps, each amounting to a 10
percent decrease in $T$. Each value of $T$ was held constant for $150$
reconfigurations, or for $15$ successful reconfigurations, whichever
was earlier.

\section{Optimized Networks: Results}

\begin{figure}[!htbp]
\centerline{
 \psfig{figure=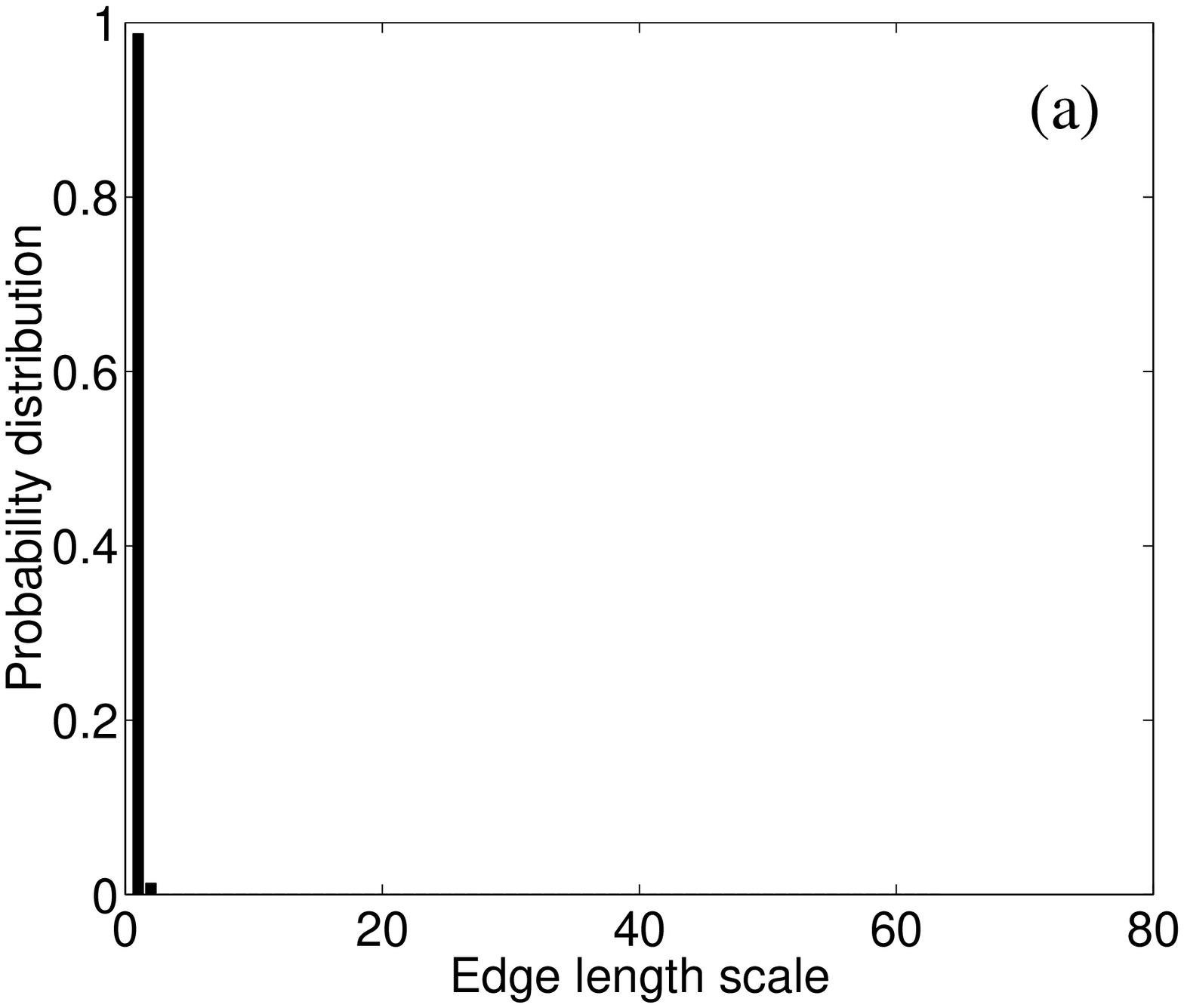,width=7.3cm,height=7.3cm} 
 \psfig{figure=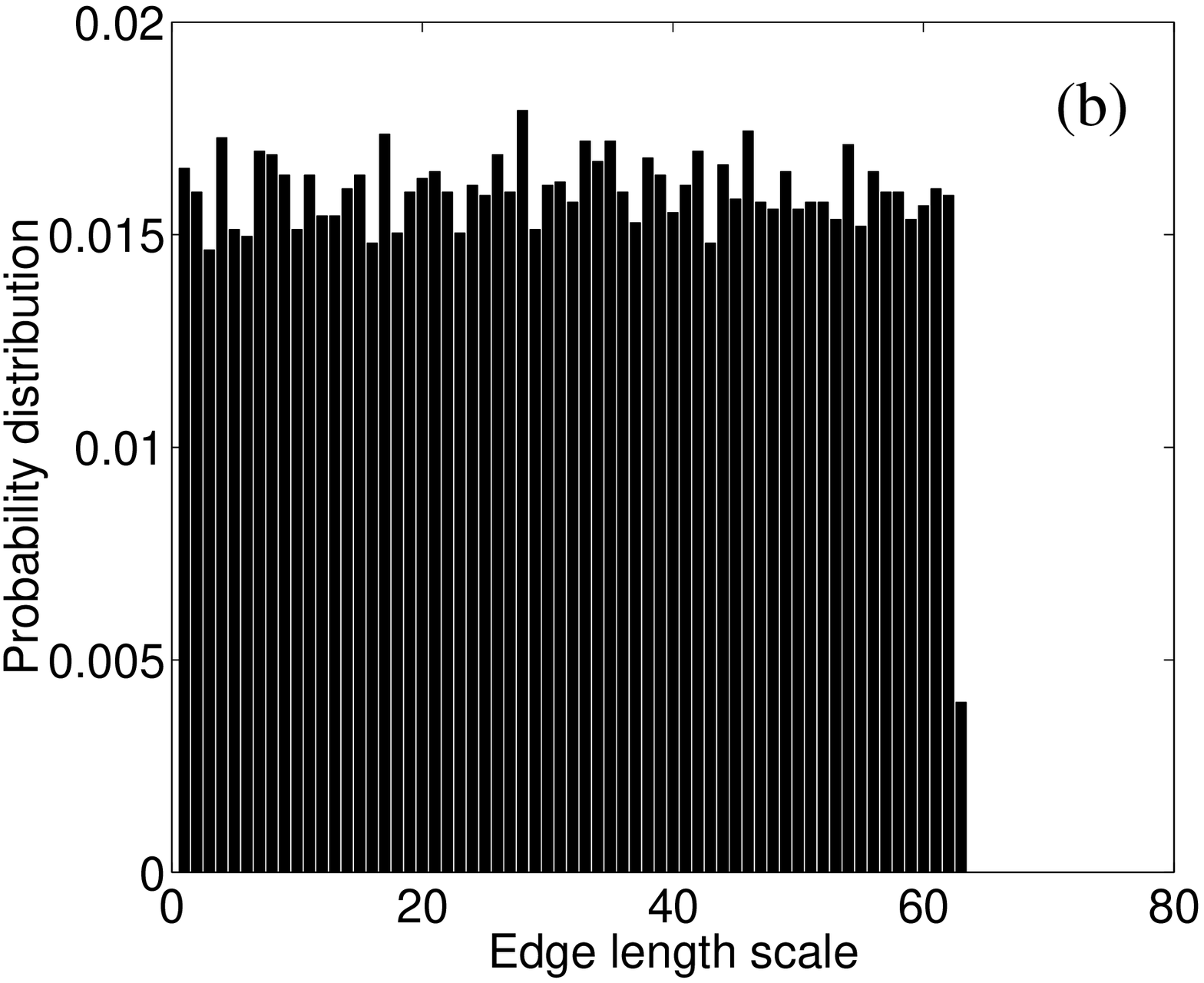,width=7.3cm,height=7.3cm}}
\caption{Edge scale distribution resulting from optimization at (a)
$\lambda=0$, and (b) $\lambda=1$ 
for a network  having $n=250$, $k=4$. Both distribution plots are averaged
over 25 simulations.} 
\label{fig:opt0-1}
\end{figure}

Since minimum characteristic path length, and minimum wiring cost are
contradictory goals, the optimization of either one or the other will
result in networks at the two ends of the randomization spectrum. As
expected, at $\lambda=0$, when the optimization function concentrates
only on minimizing the cost of wiring edges, a regular network emerges with
uniform connectivity and high characteristic path length ($L \sim n$).
The edge scale distribution shows all edges to be concentrated almost
entirely within the unit length scale, as shown in
Fig.\ \ref{fig:opt0-1} (a). At $\lambda=1$, when only the
characteristic path length is to be minimized, again of no surprise,
the optimization results in a near random network ($L \sim \ln n$). The
edge scale distribution shown in Fig.\ \ref{fig:opt0-1} (b) has edges
having lengths distributed uniformly over the entire length scale
range.

\begin{figure}[!htbp]
\centerline{
   \psfig{figure=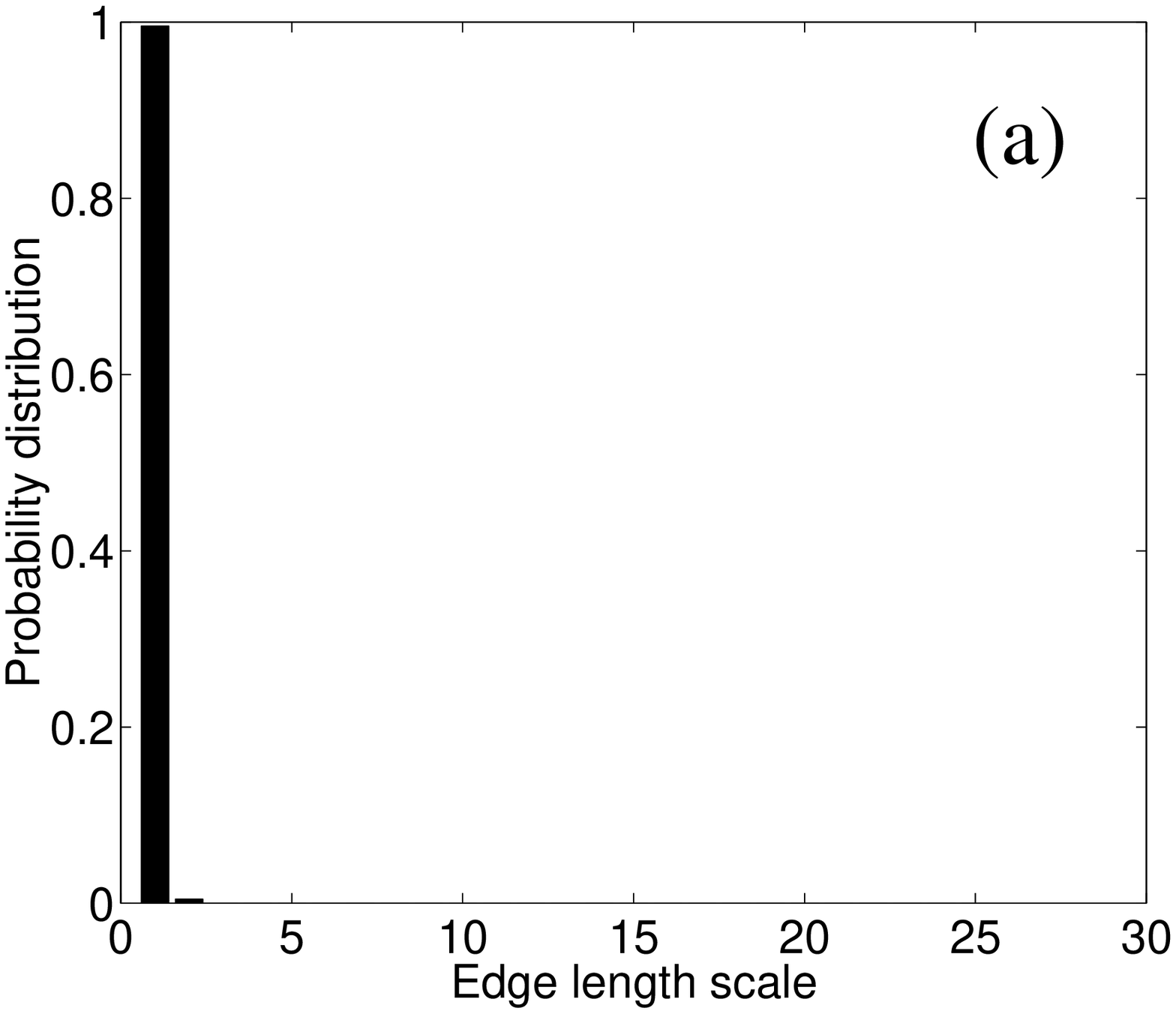,width=5.0cm,height=5.0cm}
   \psfig{figure=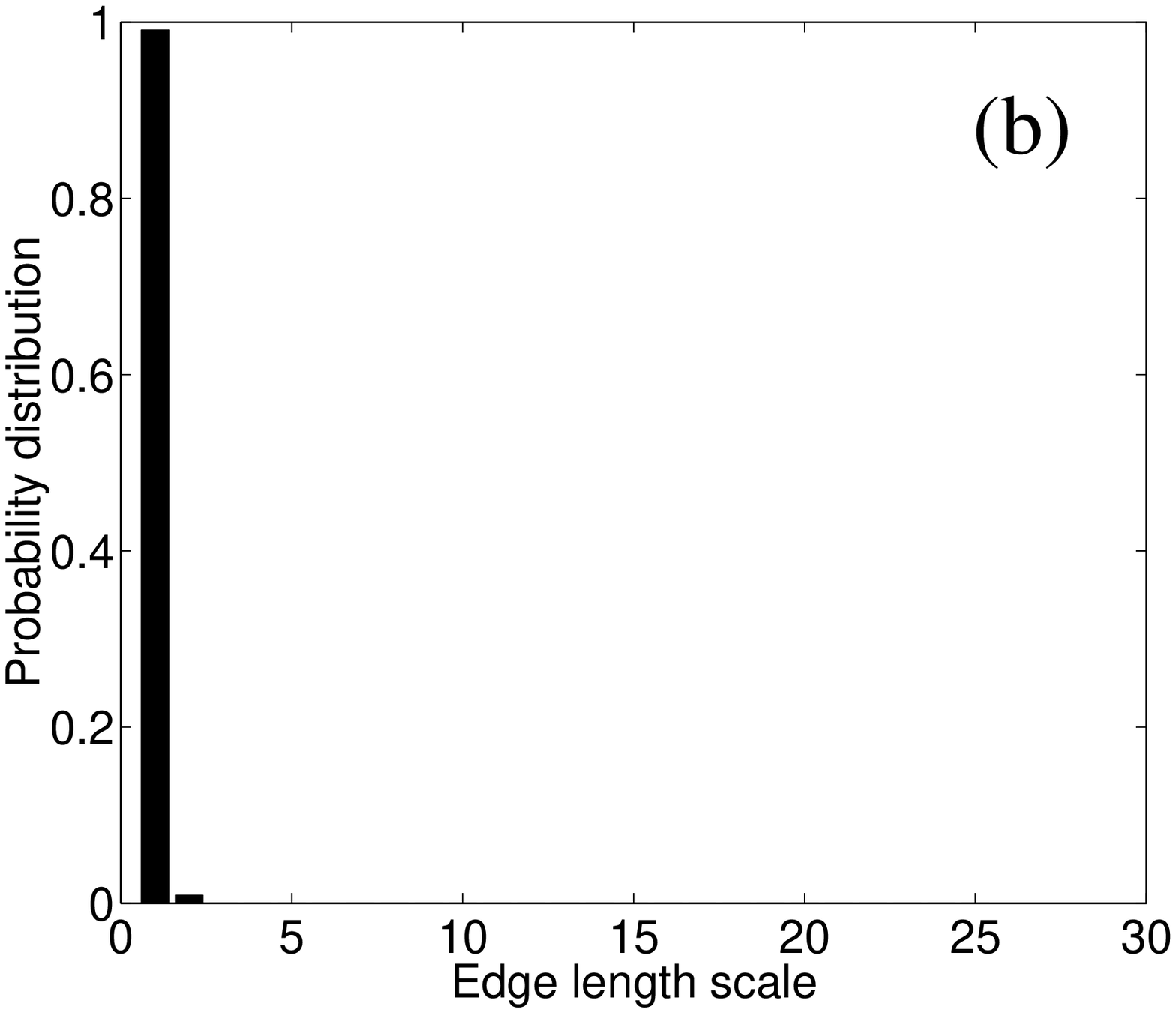,width=5.0cm,height=5.0cm}
   \psfig{figure=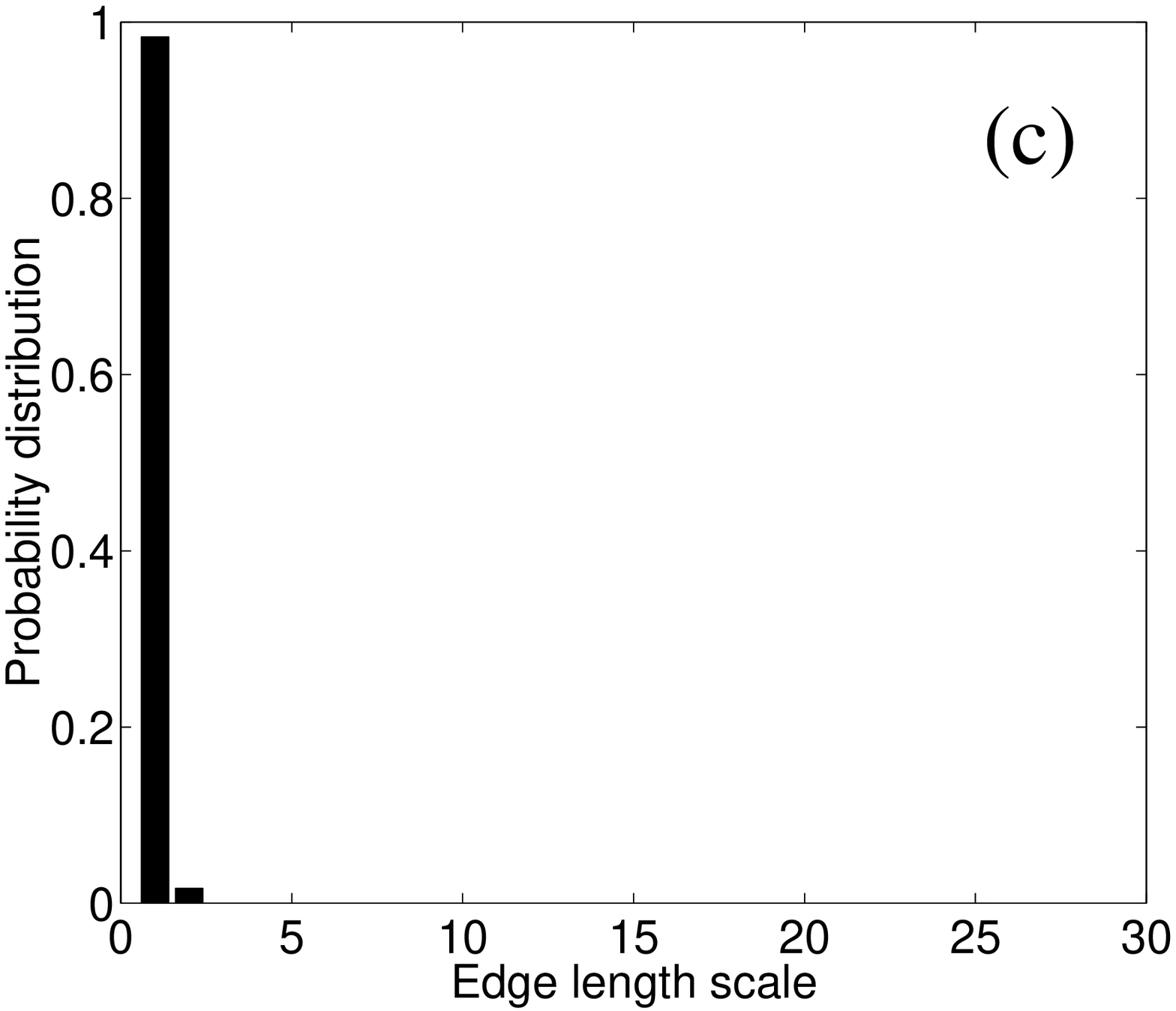,width=5.0cm,height=5.0cm}}
\centerline{
   \psfig{figure=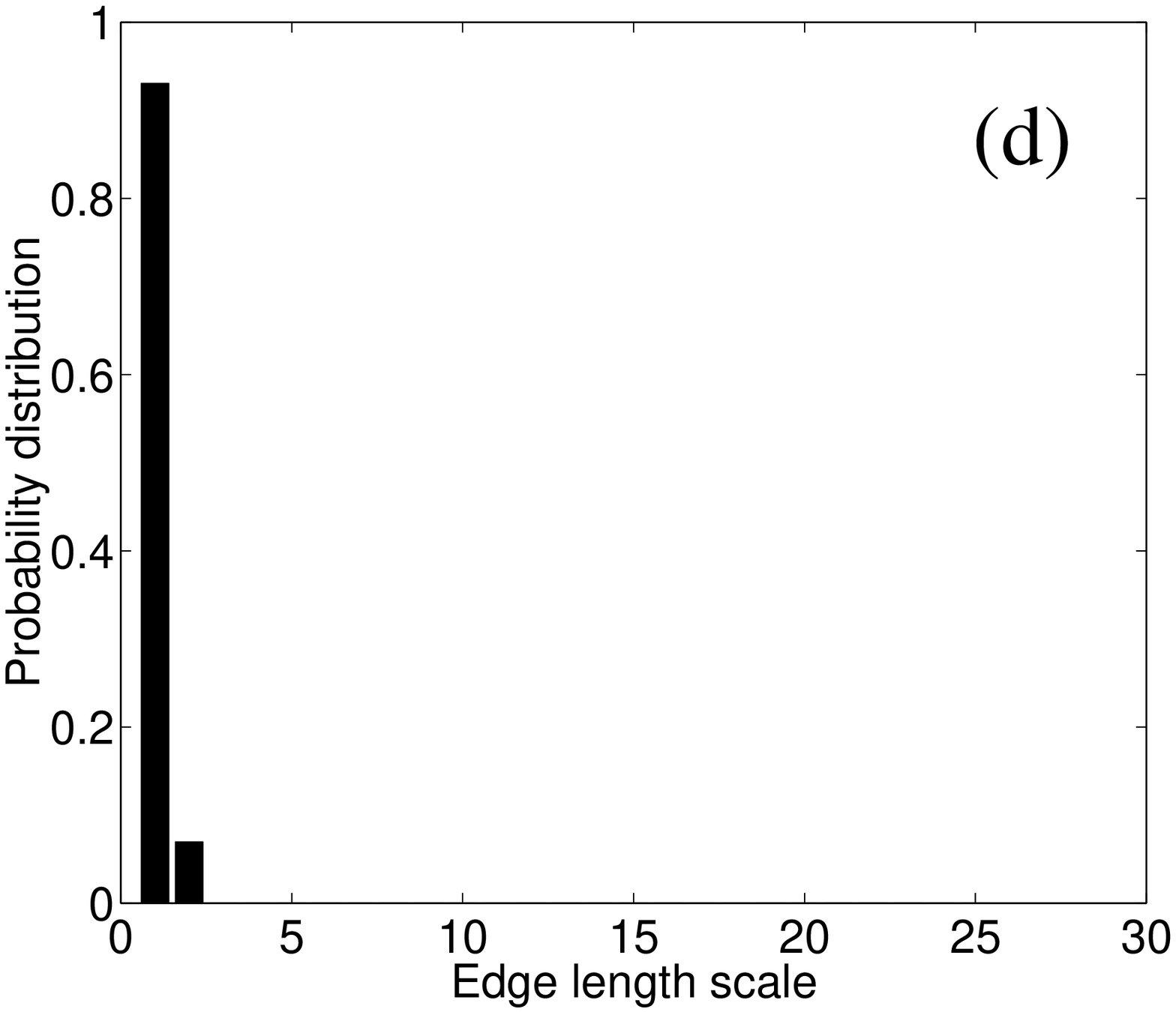,width=5.0cm,height=5.0cm}
   \psfig{figure=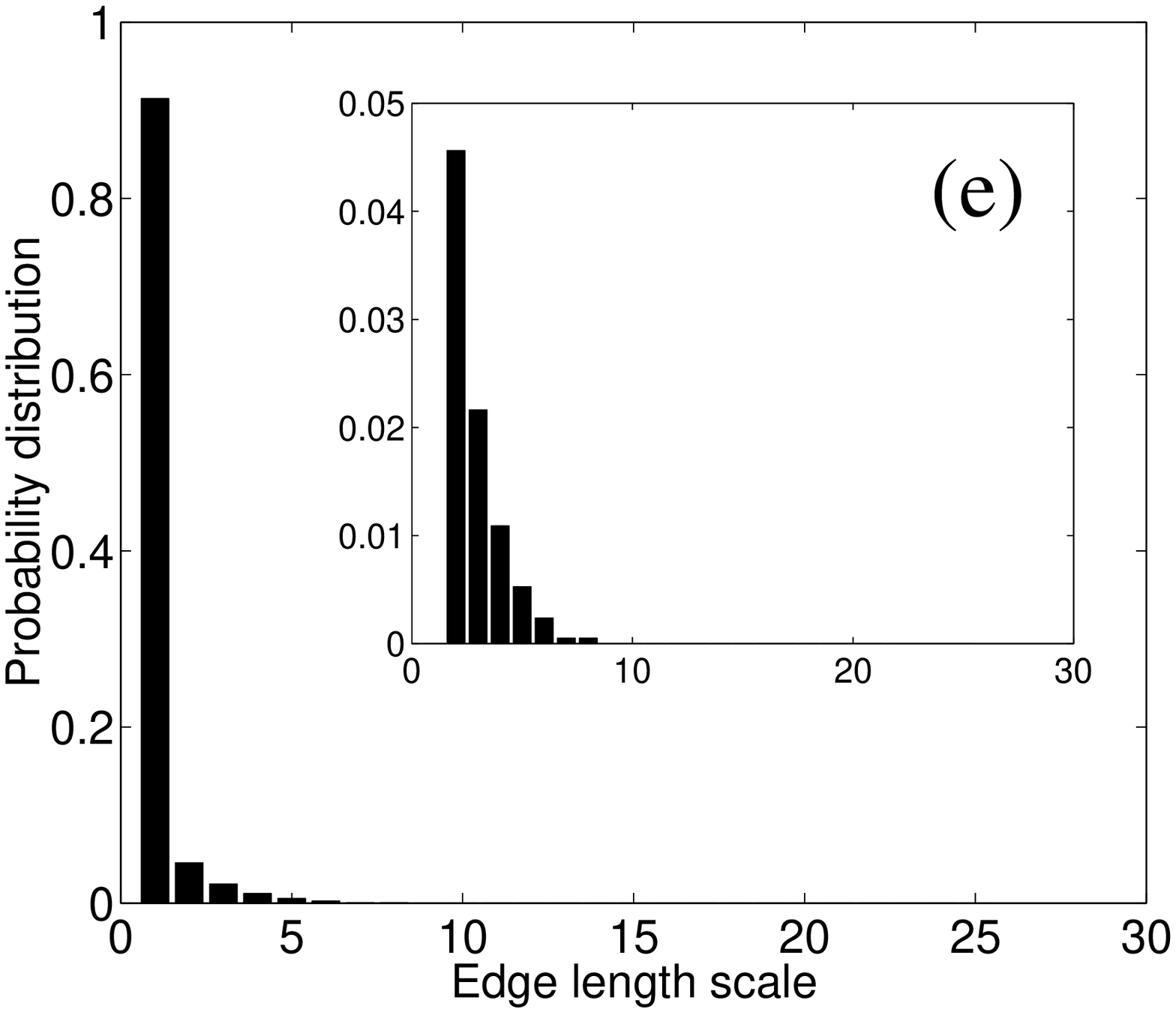,width=5.0cm,height=5.0cm}
   \psfig{figure=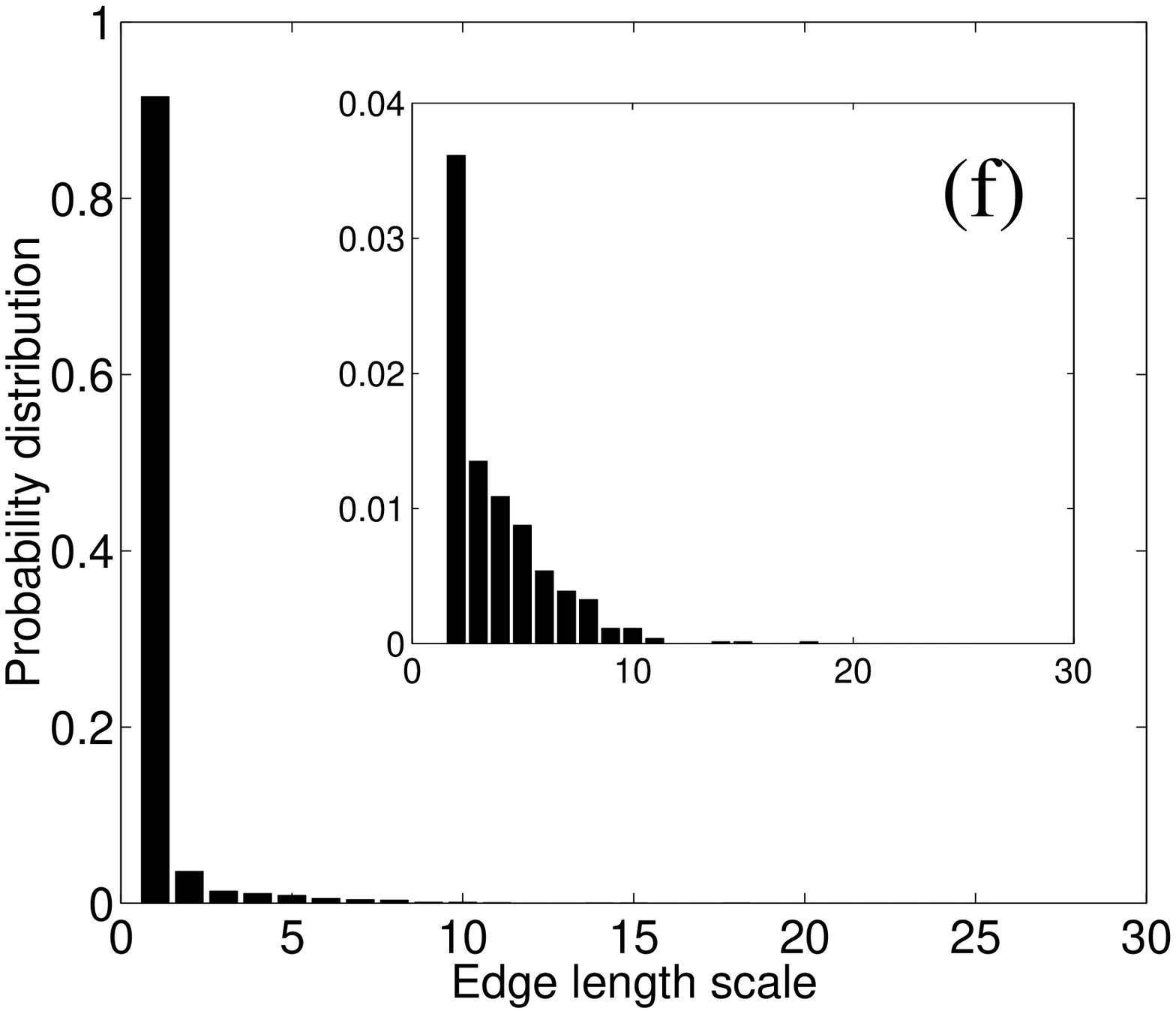,width=5.0cm,height=5.0cm}}    \centerline{
   \psfig{figure=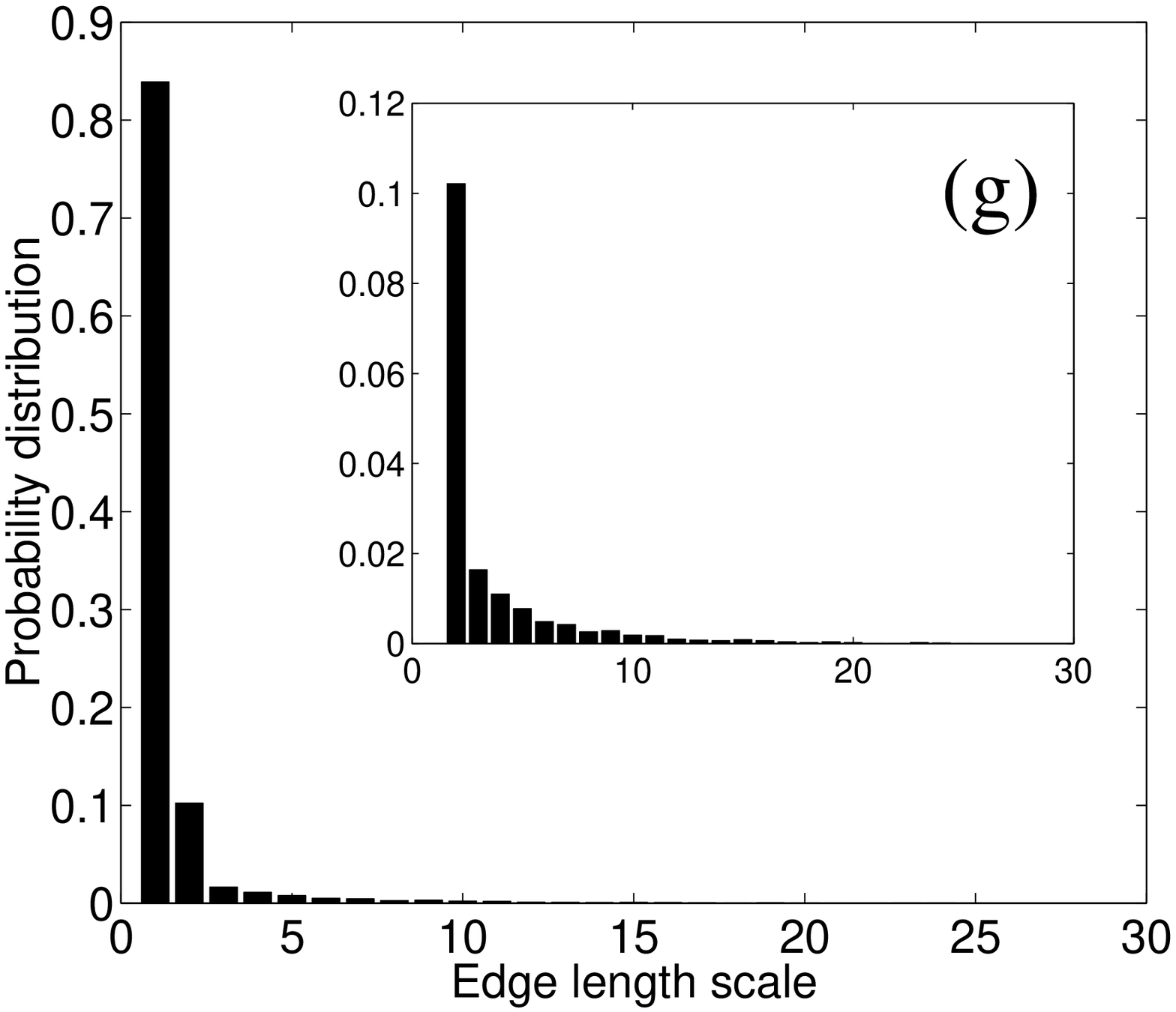,width=5.0cm,height=5.0cm}
   \psfig{figure=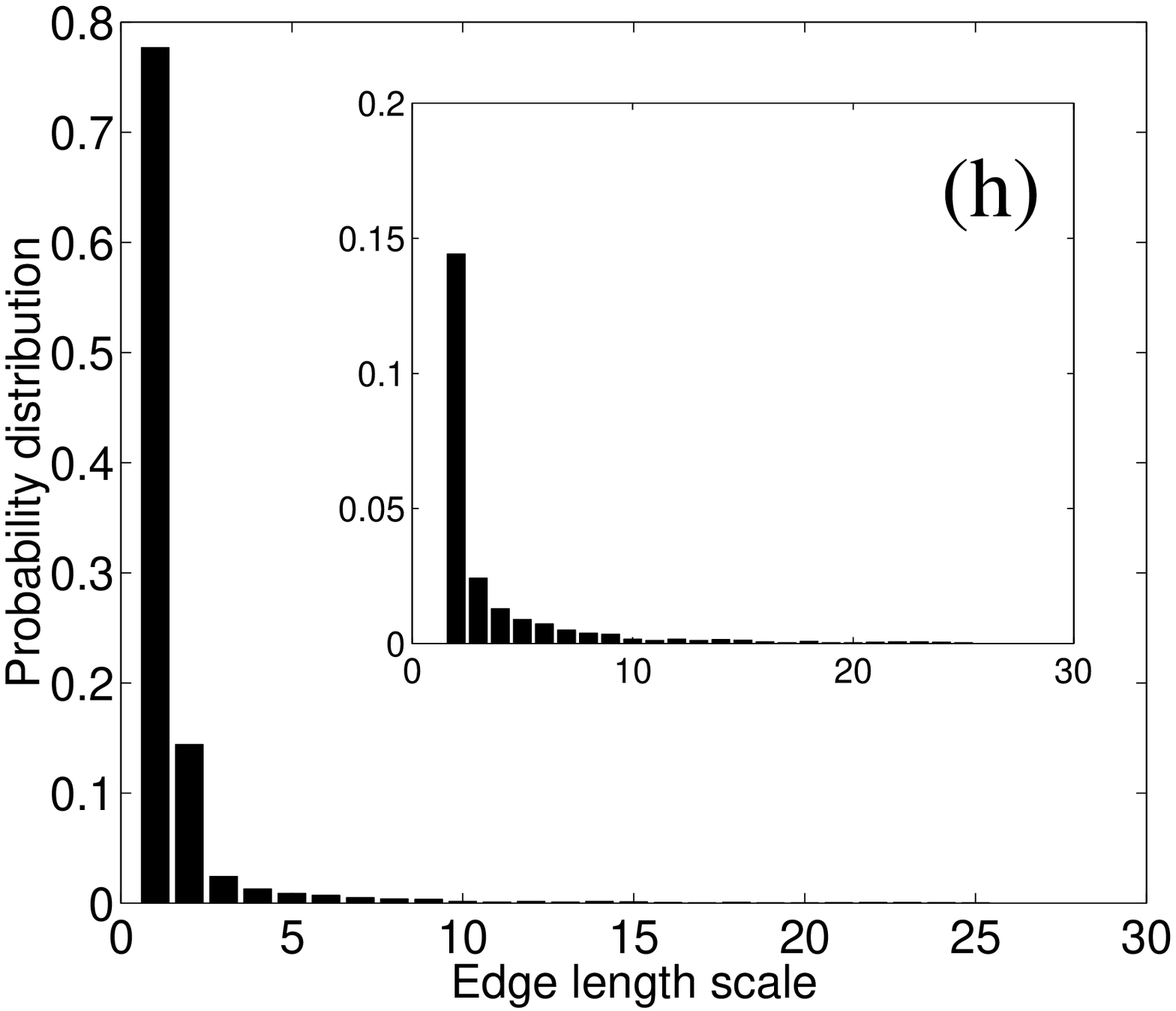,width=5.0cm,height=5.0cm}
   \psfig{figure=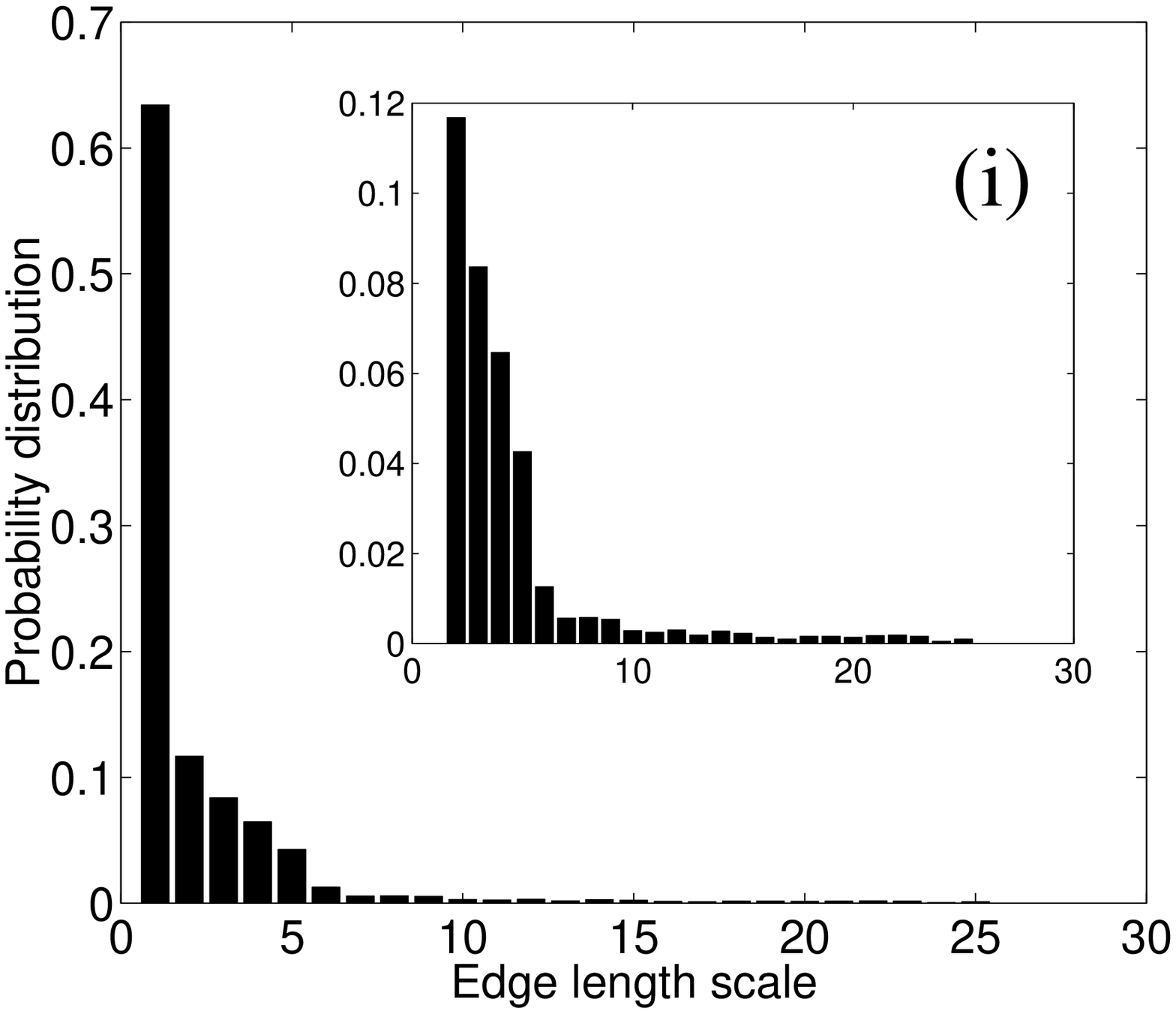,width=5.0cm,height=5.0cm}}
\centerline{
   \psfig{figure=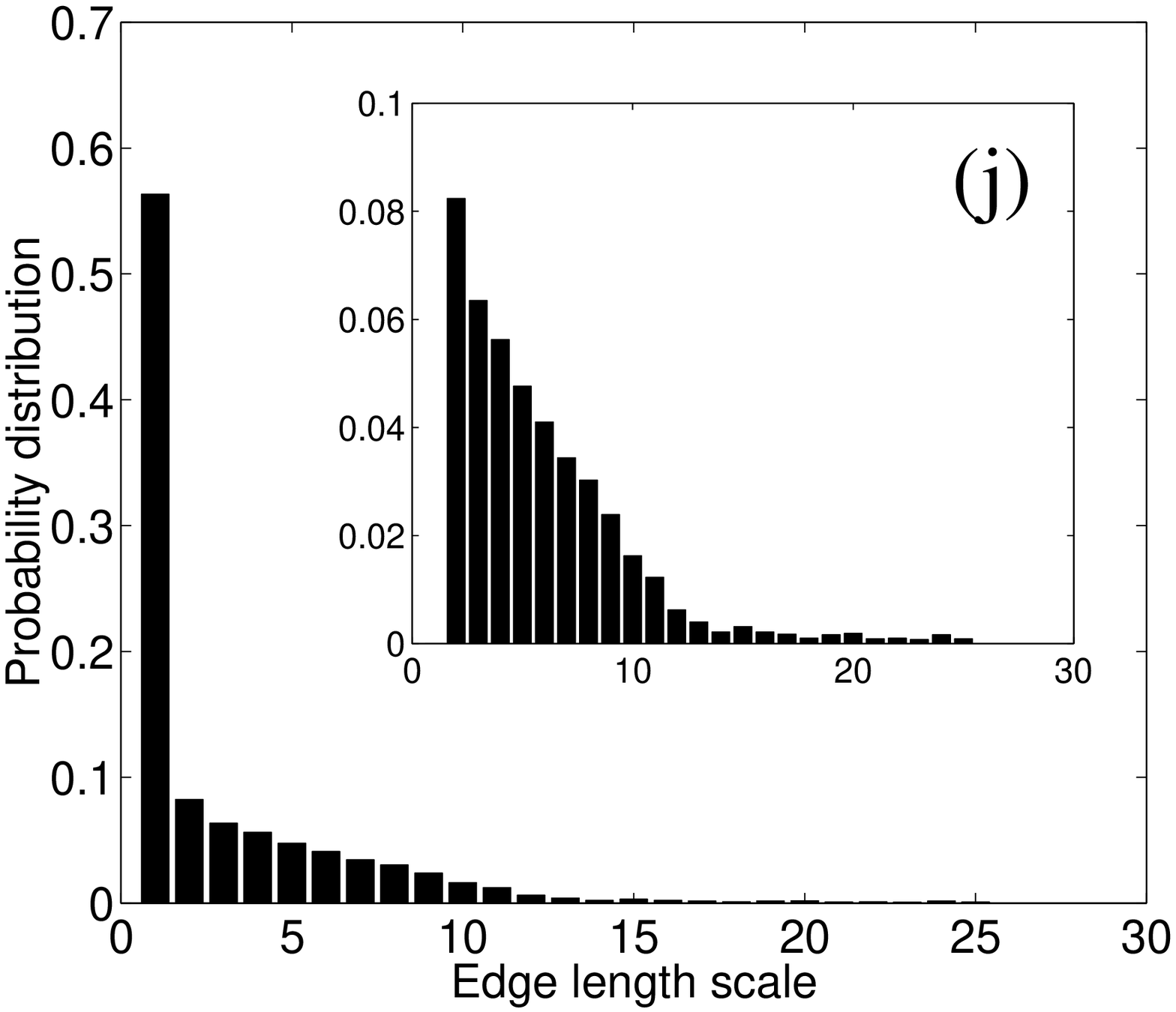,width=5.0cm,height=5.0cm}
   \psfig{figure=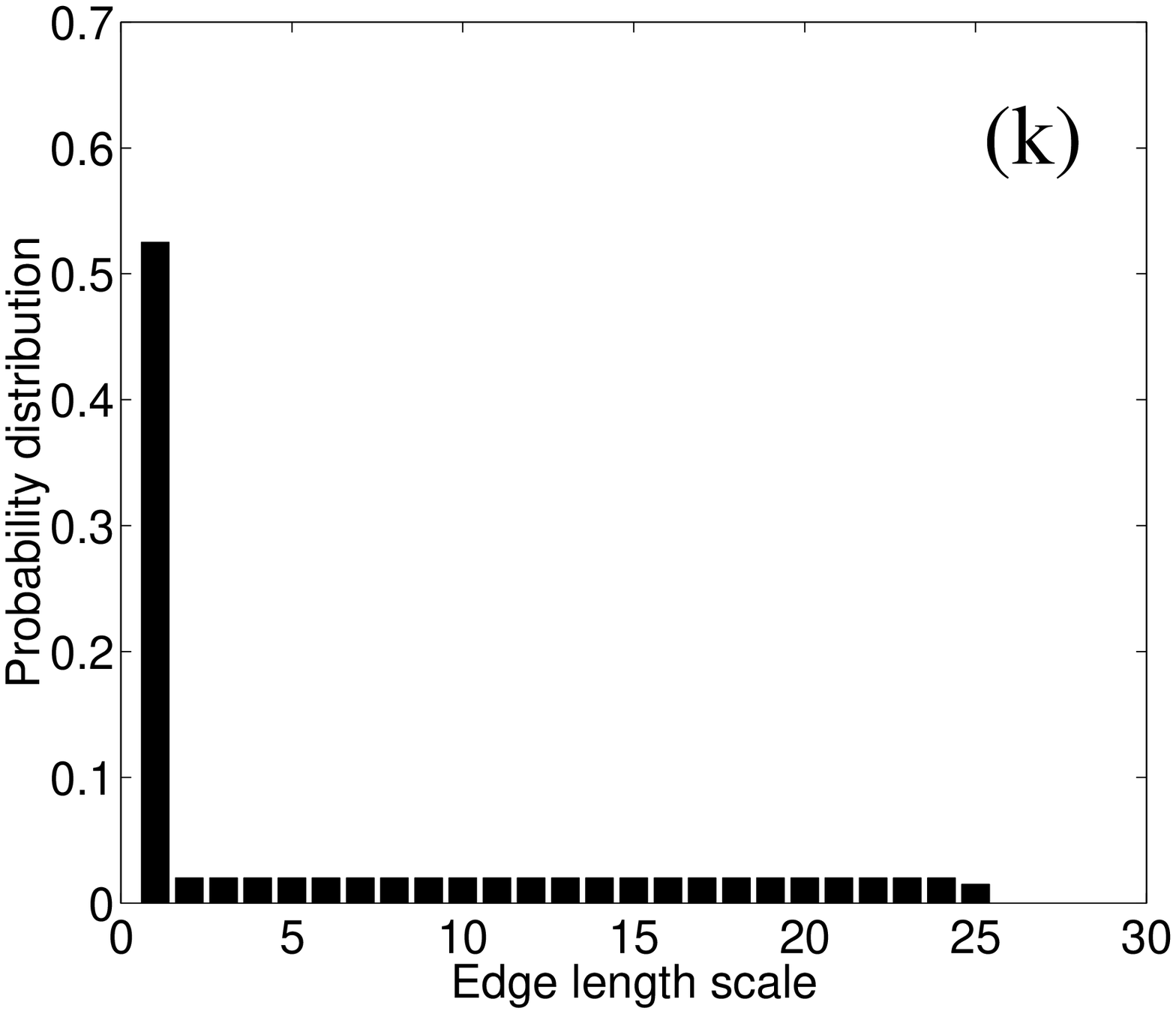,width=5.0cm,height=5.0cm}
   \psfig{figure=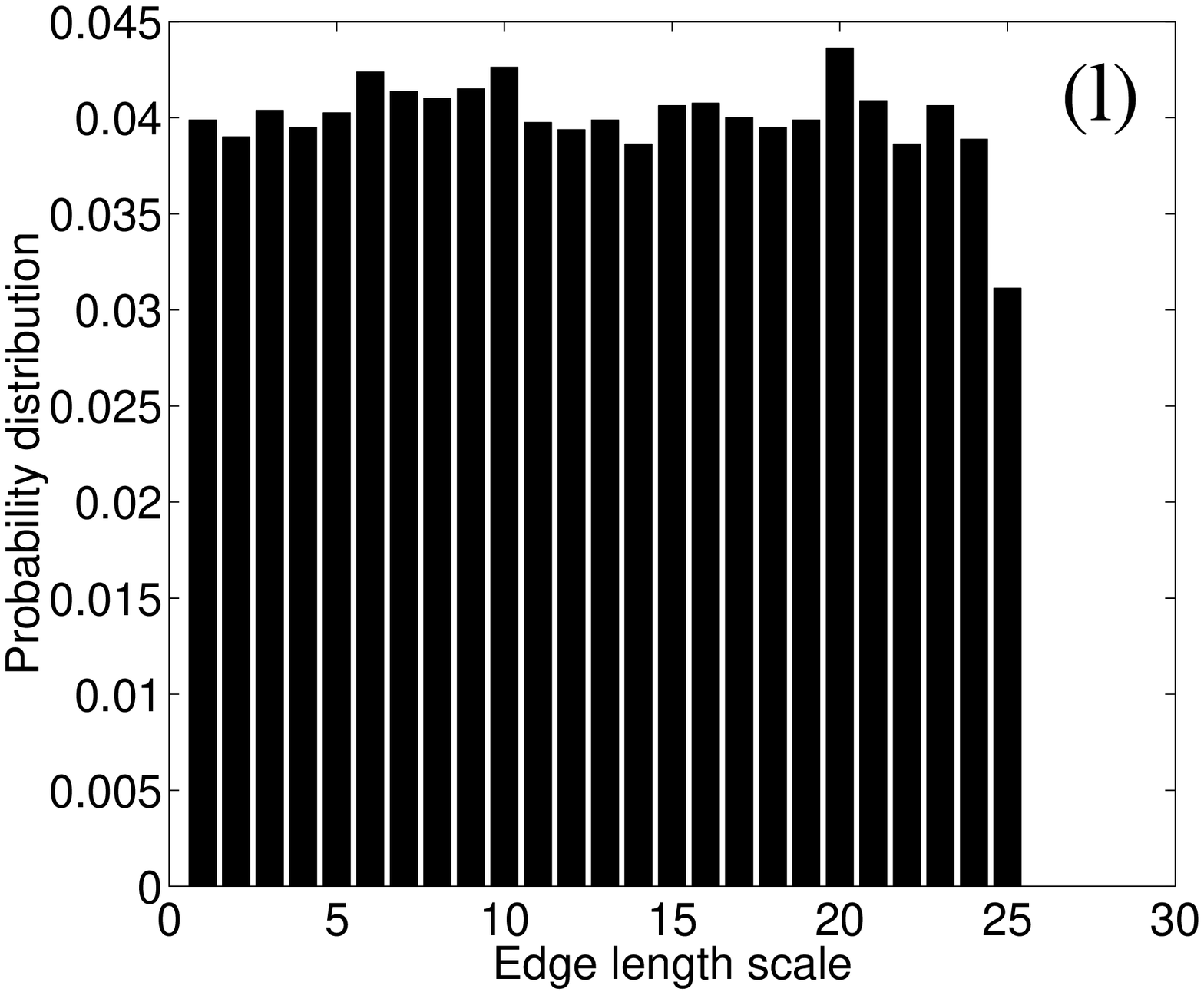,width=5.0cm,height=5.0cm}}
\caption{Edge scale distribution for an $n$=100, $k$=4 network at various
$\lambda$: 
(a) 0.0, (b) $5 \times 10^{-4}$, (c) $5 \times 10^{-3}$, 
(d) $1.25 \times 10^{-2}$, (e) $2.5 \times 10^{-2}$, (f) $5 \times 10^{-2}$, 
(g) $1.25 \times 10^{-1}$, (h) $2.5 \times 10^{-1}$, (i) $5 \times 10^{-1}$,
(j) $7.5 \times 10^{-1}$, (k) $8.5 \times 10^{-1}$, (l) 1.0. 
The inset in each plot shows the distribution of all scales with the unit 
length scale excluded. Each plot is an average over 40 simulations.} 
\label{fig:optEdgeScaleDist}
\end{figure}  

\subsection{The emergence of hubs}

At intermediate values of $\lambda$, the optimization model results in
{\em hubs}, that is, a group of nodes connected to a single node. Due
to the constraint which seeks to minimize physical distance between
connected vertices, hubs are formed by vertices close to one another.
In addition, the minimization of graph distance ensures the existence
of connections between hub centres, enabling whole hubs to communicate
with each other. The edges at any hub centre therefore, span a wide
range of length scales. Hubs emerge due to the
contribution of $L$ to the optimization function. The formation of
hubs, en route to the emergence of a small-world network, has so far not been reported in the literature.

The extreme situation is a `universal' hub:  a single node,  
with all other nodes having connections to it. However,
except for the situations when the cost of wiring is negligible, 
 we find that the optimization does not result in a
universal hub. This is apparent, since a universal hub requires all
the remaining $n-1$ vertices to have connections to the vertex at the
centre of the hub, resulting in length scales which span the entire
scale range, long connections being prohibitively expensive. A real
world example of such a universal hub network is unlikely since a
large hub is a bottleneck to traffic through it, resulting in
overcrowding at the hubs \cite{3:kasturi}. Hence, the need for
multiple, and consequently smaller, hubs.

Watts \cite{3:watts2} defines the {\em significance} of a vertex $v$, as the 
characteristic path length of its neighbourhood $\Gamma(v)$, in its absence. 
Hub centres are significant since they
contract distances {\em between} every pair of vertices within the
hub. Thus, vertex pairs although not directly connected, are
connected via the single common vertex. Hence, the average
significance, a measure which reflects the number of contractions, is
considerable. Thus, in contrast to the WS model, where networks
become small due to shortcuts, here smallness can be attributed to the
small fraction of highly significant vertices.

\begin{figure}[!htbp]
\centerline{\psfig{figure=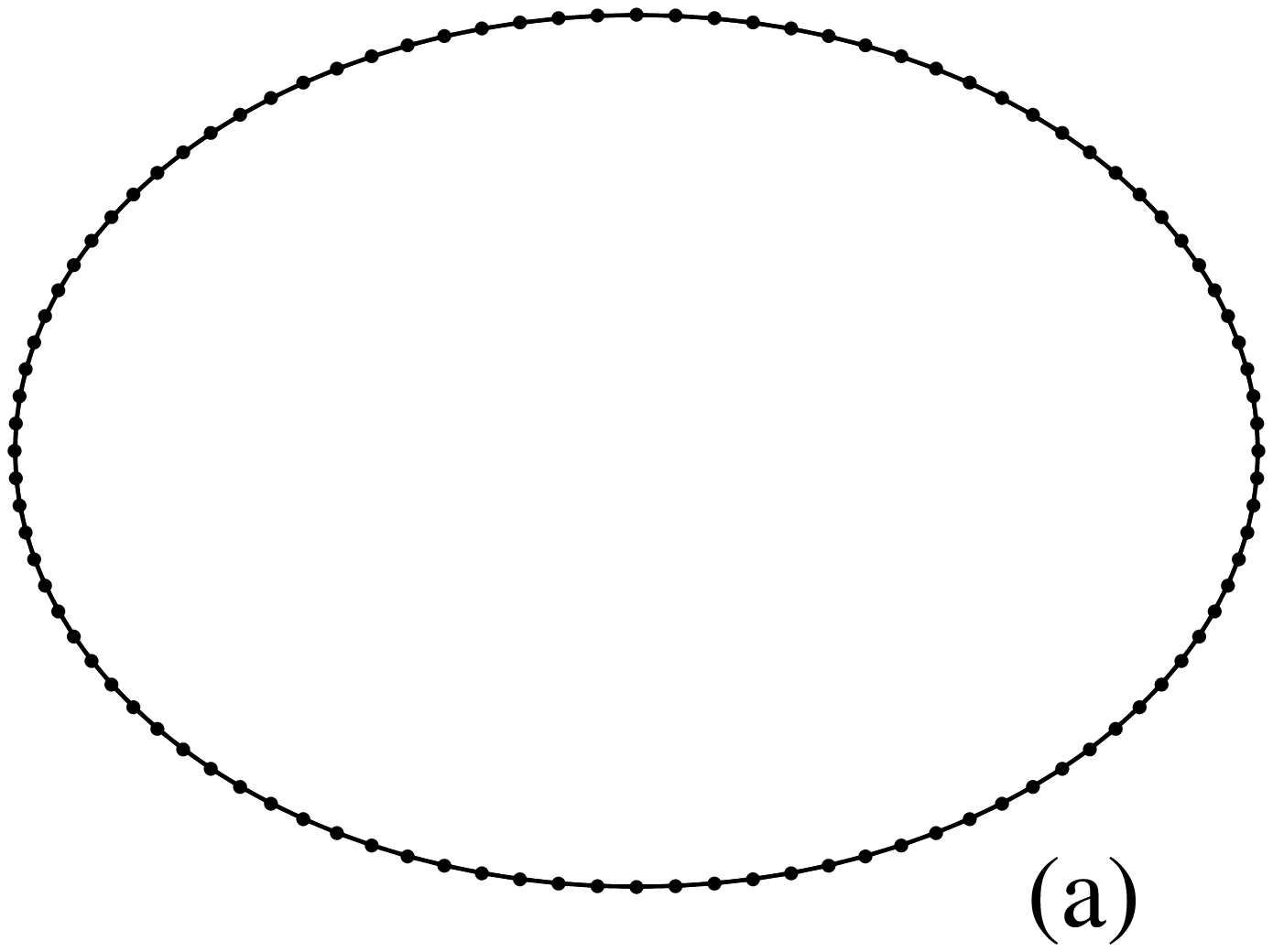,width=4.8cm,height=4.8cm}
            \psfig{figure=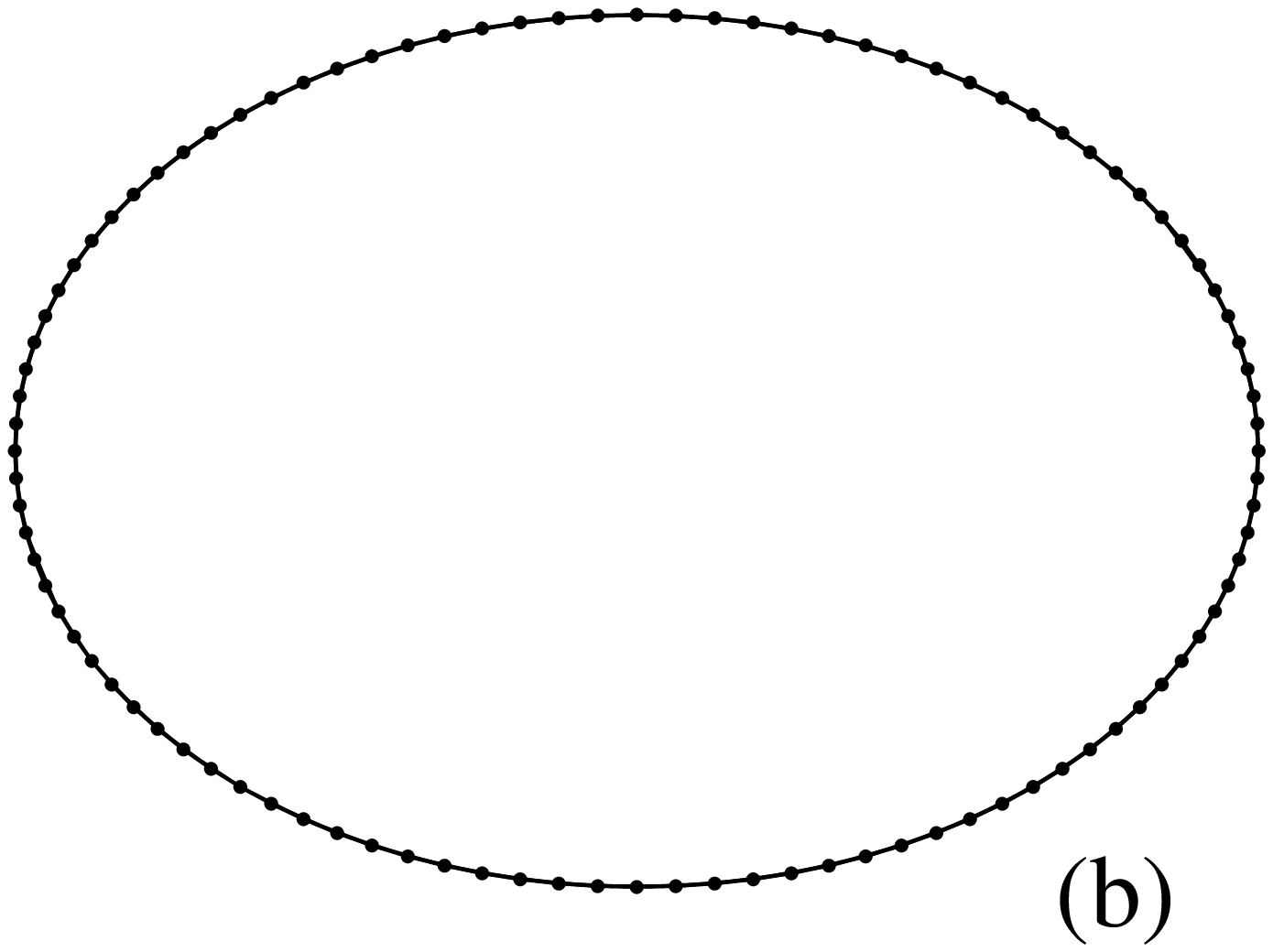,width=4.8cm,height=4.8cm}
            \psfig{figure=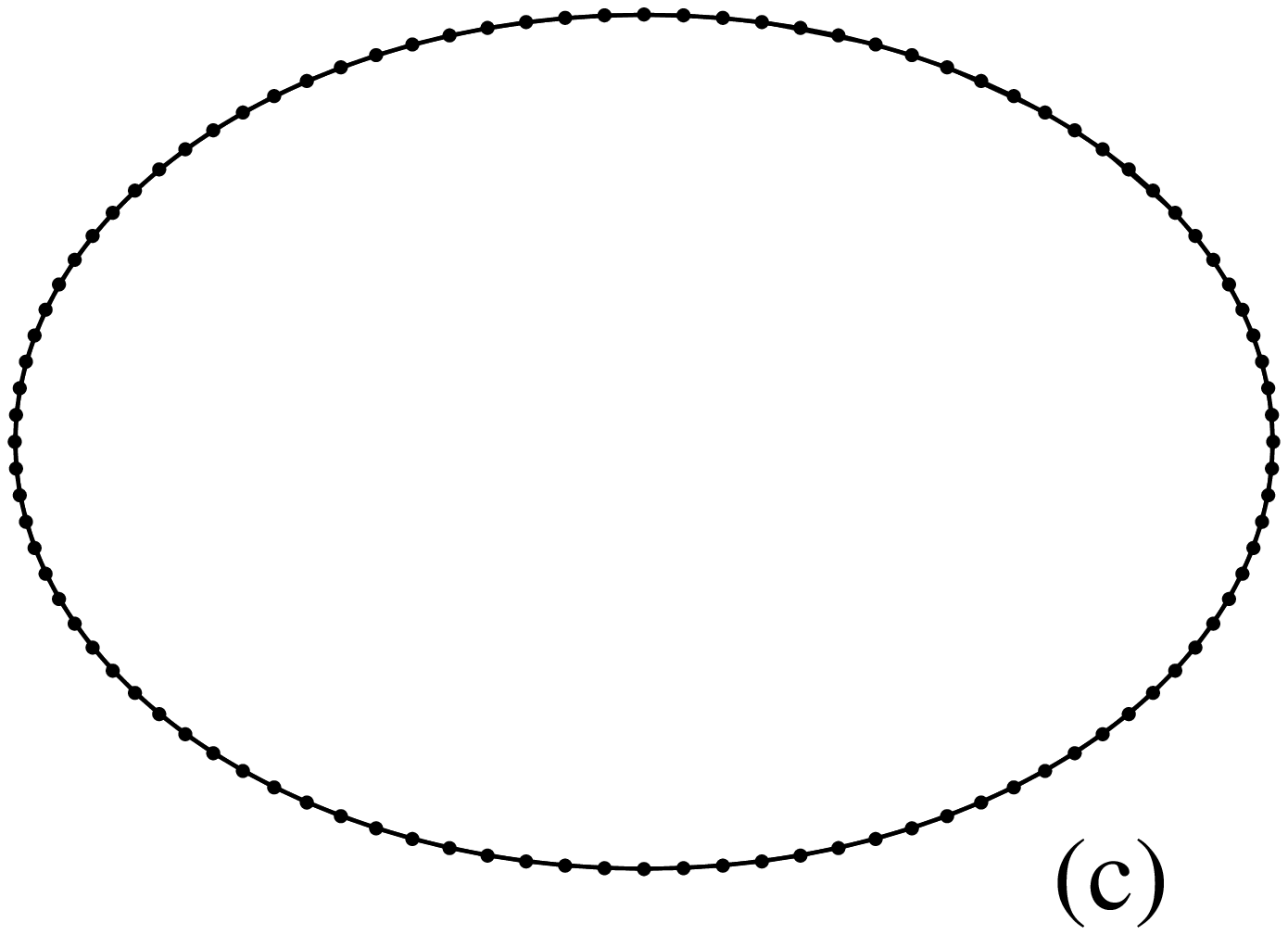,width=4.8cm,height=4.8cm}}
\centerline{\psfig{figure=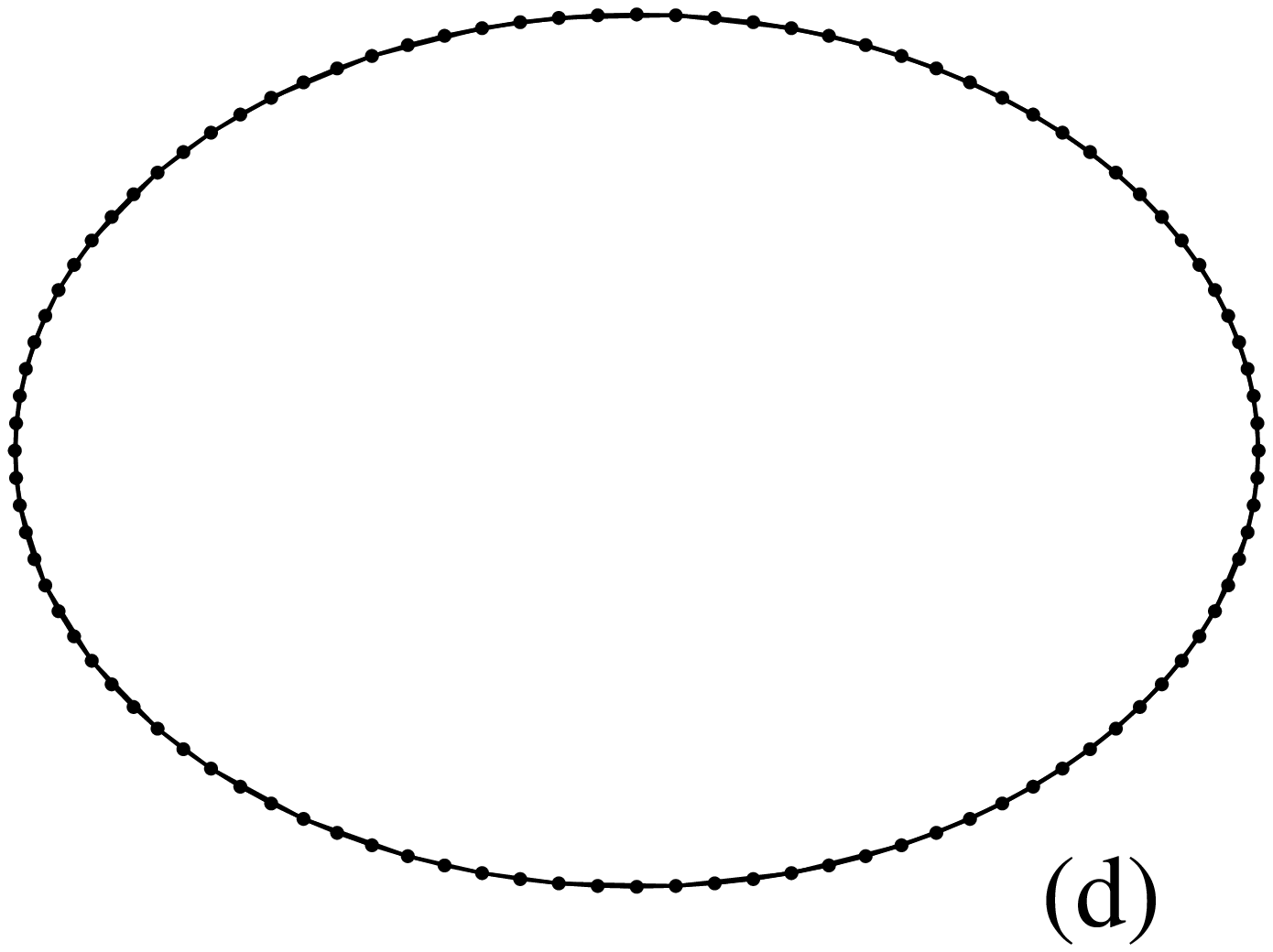,width=4.8cm,height=4.8cm}
            \psfig{figure=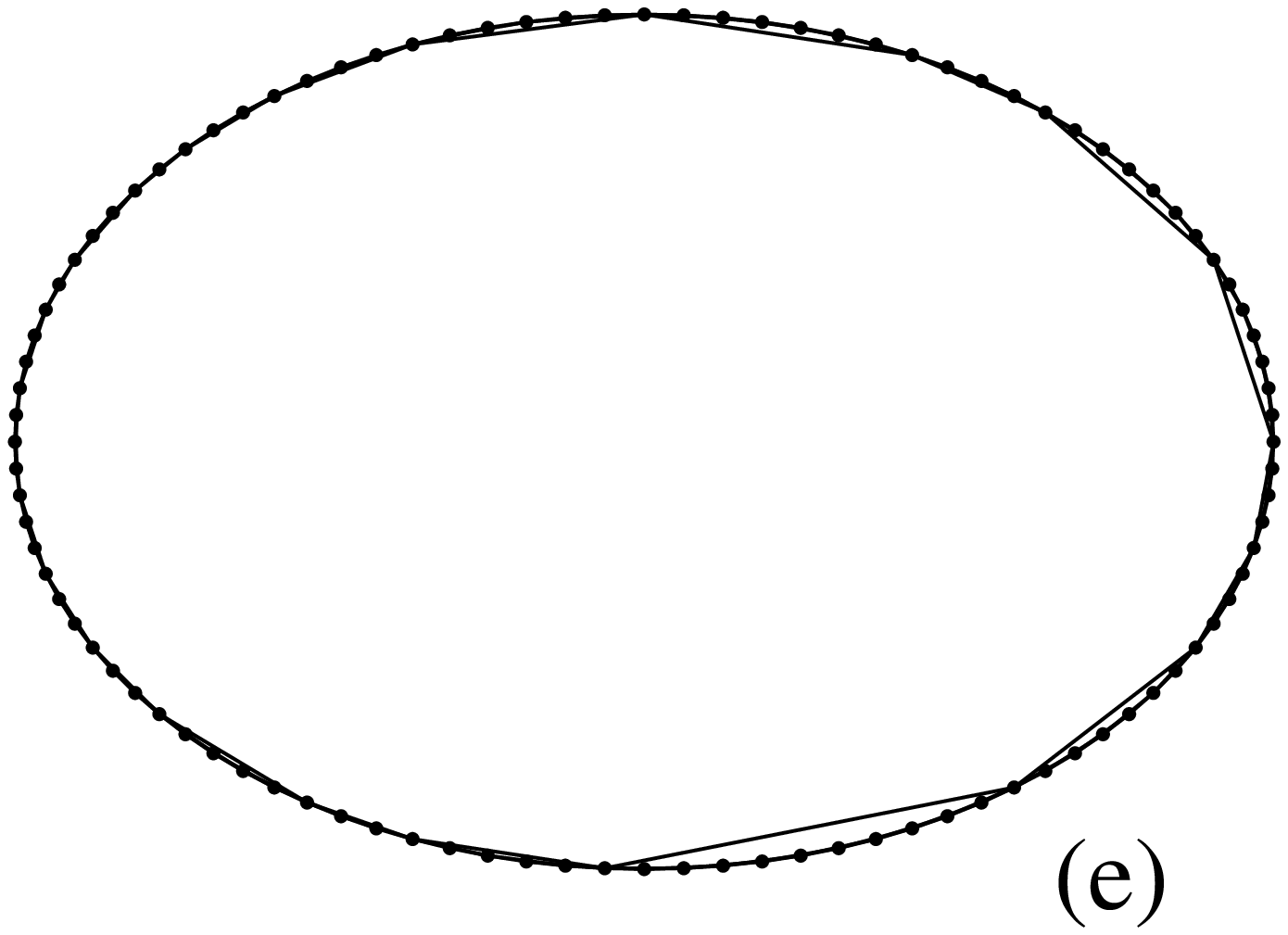,width=4.8cm,height=4.8cm}
            \psfig{figure=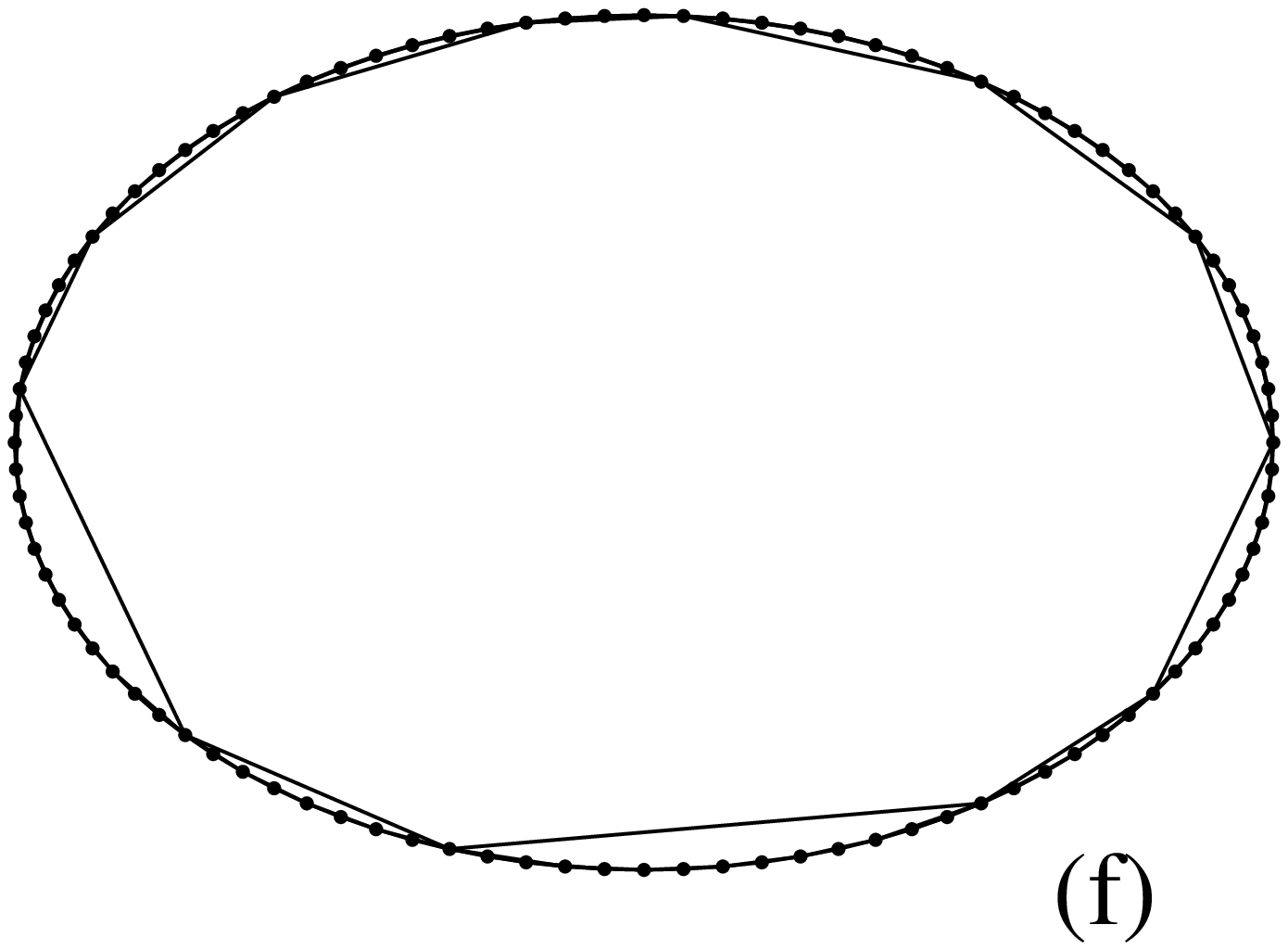,width=4.8cm,height=4.8cm}}            
\centerline{\psfig{figure=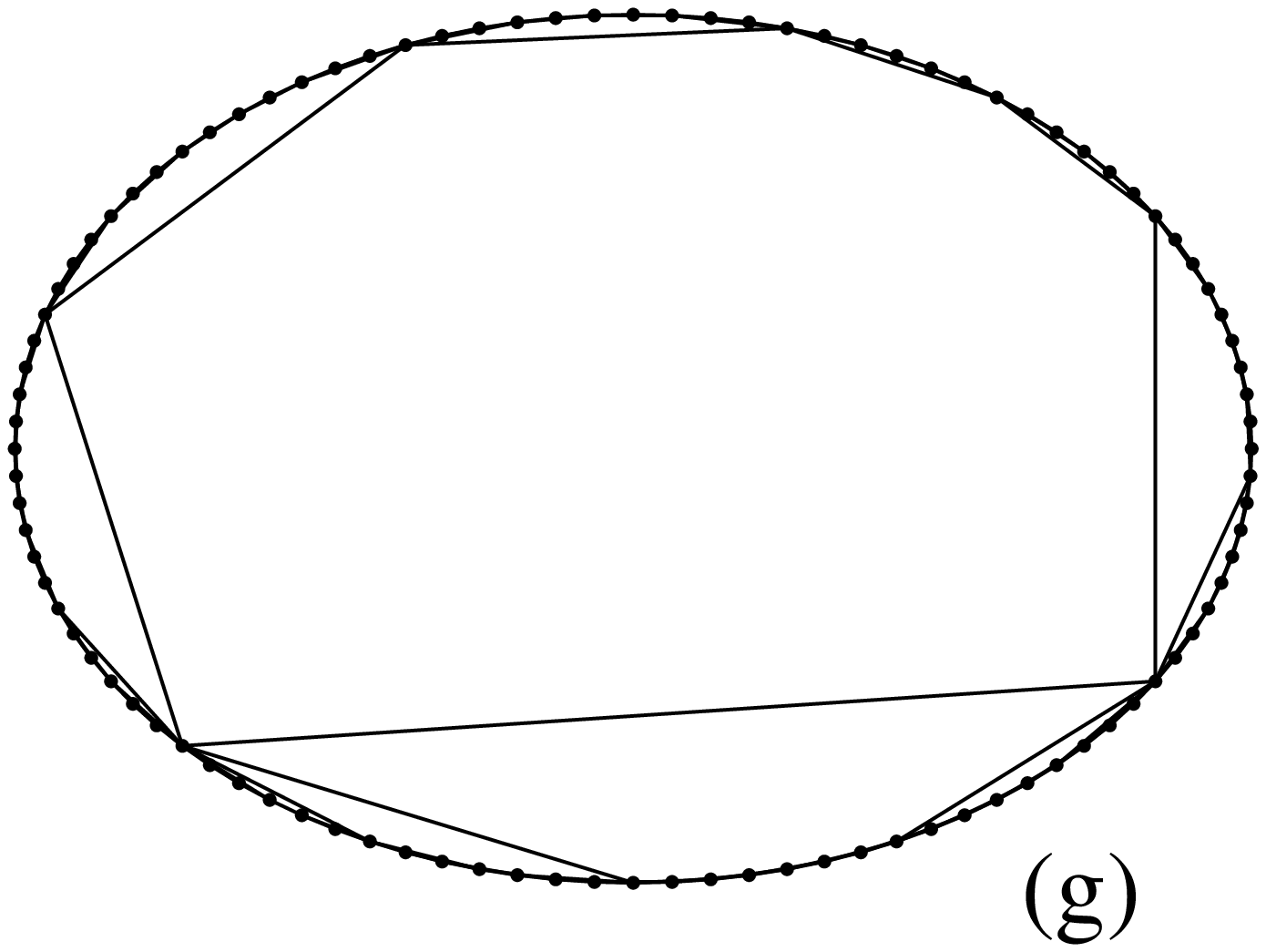,width=4.8cm,height=4.8cm}
            \psfig{figure=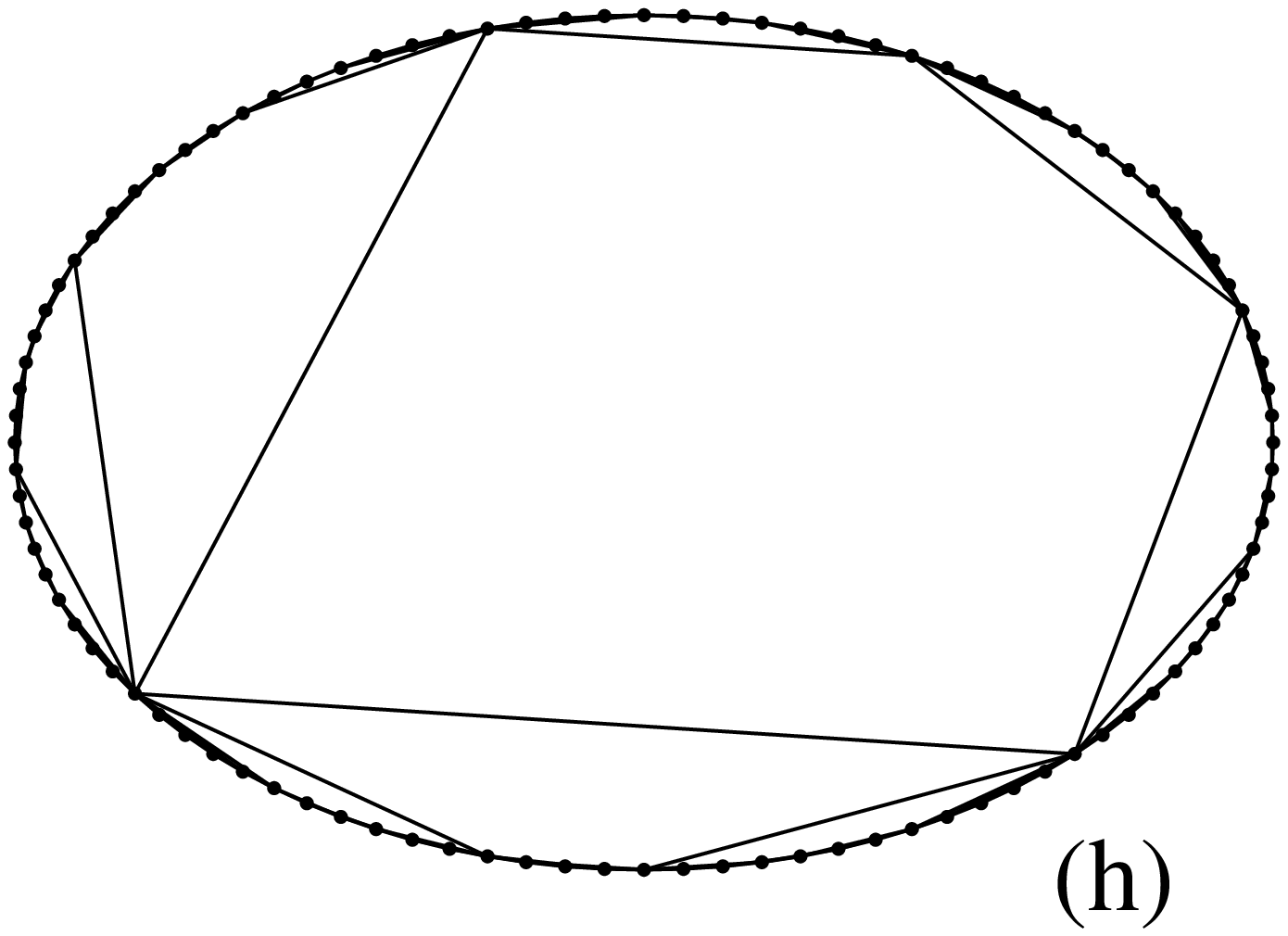,width=4.8cm,height=4.8cm}
            \psfig{figure=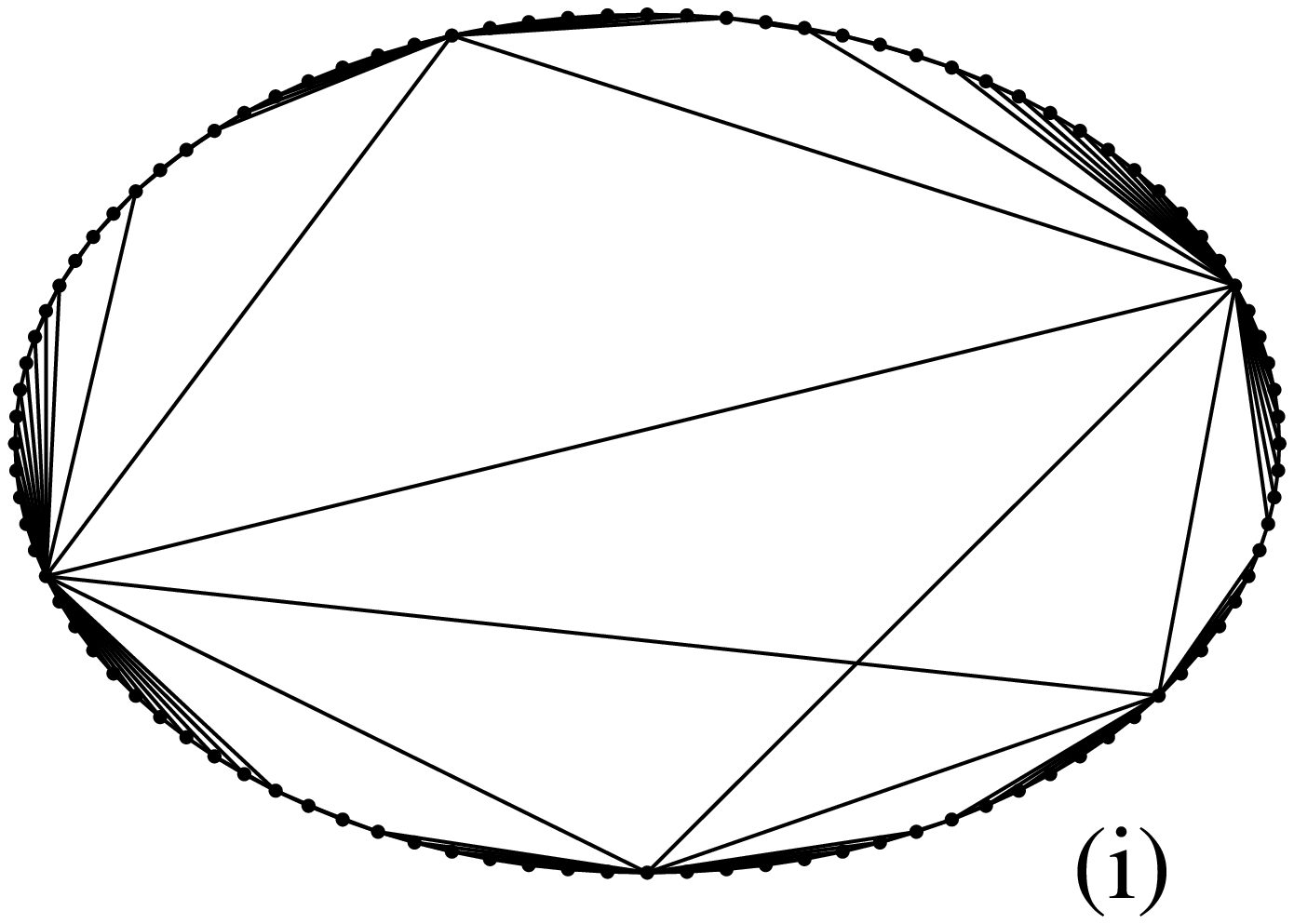,width=4.8cm,height=4.8cm}}
\centerline{\psfig{figure=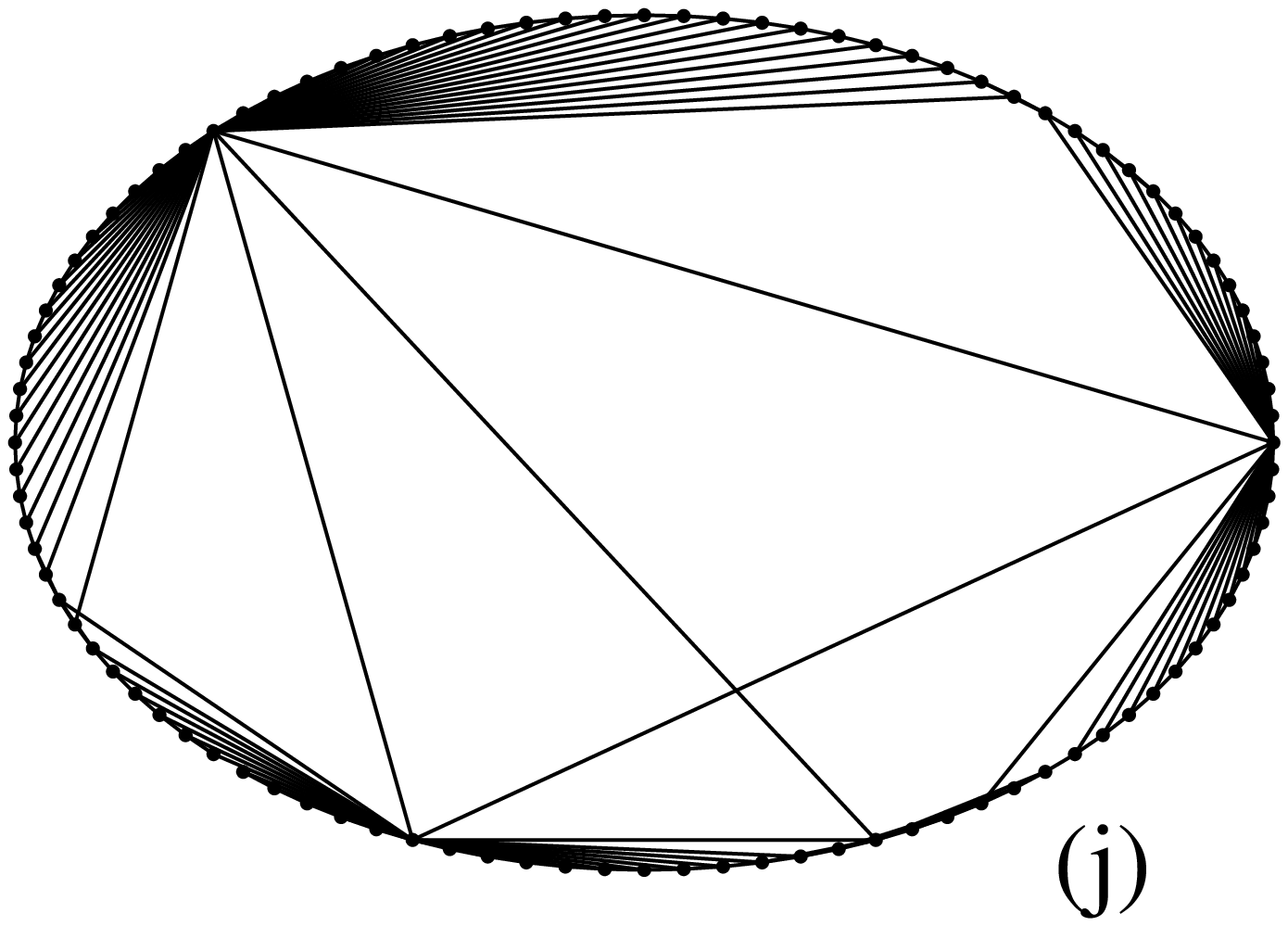,width=4.8cm,height=4.8cm}
            \psfig{figure=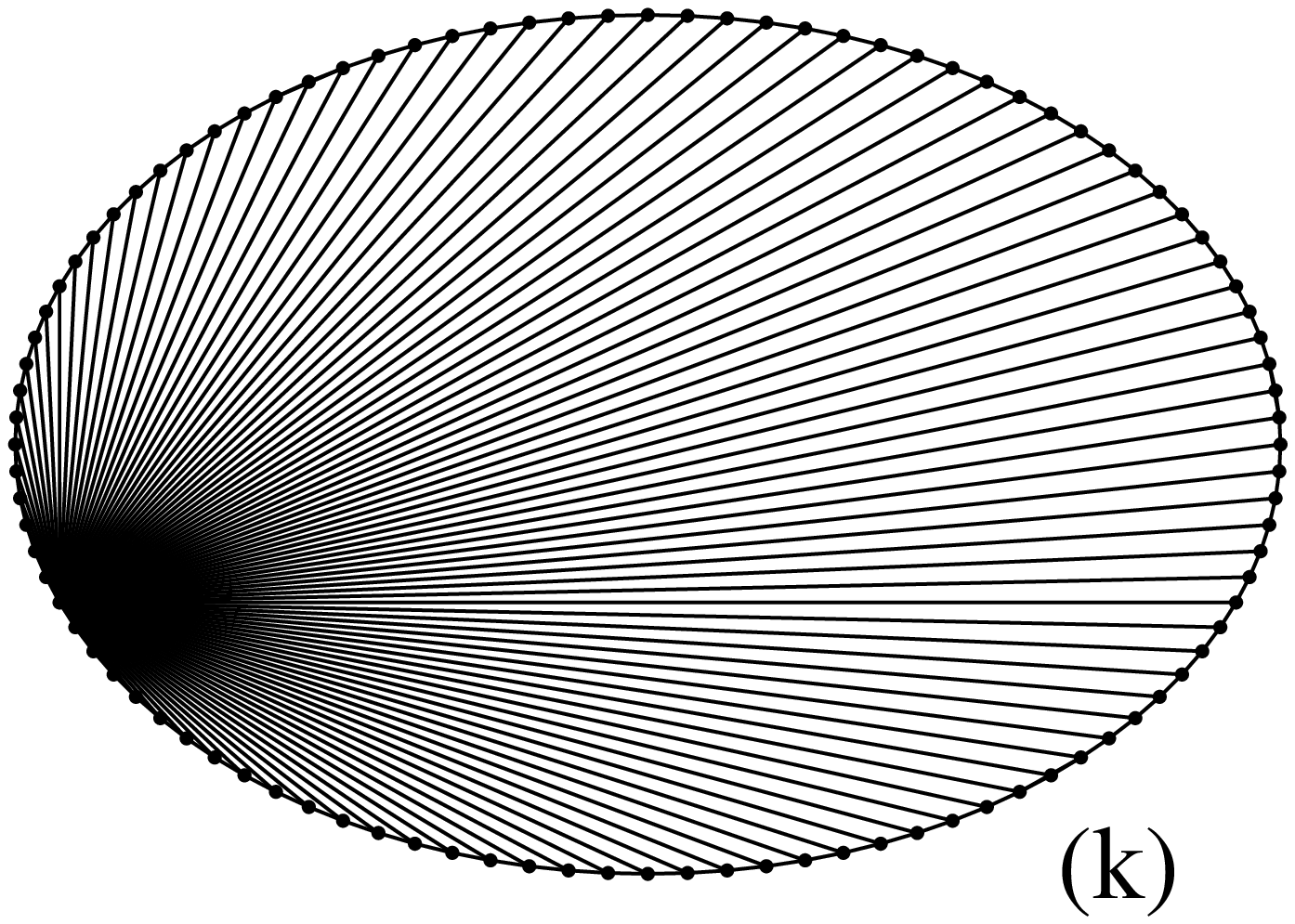,width=4.8cm,height=4.8cm}
            \psfig{figure=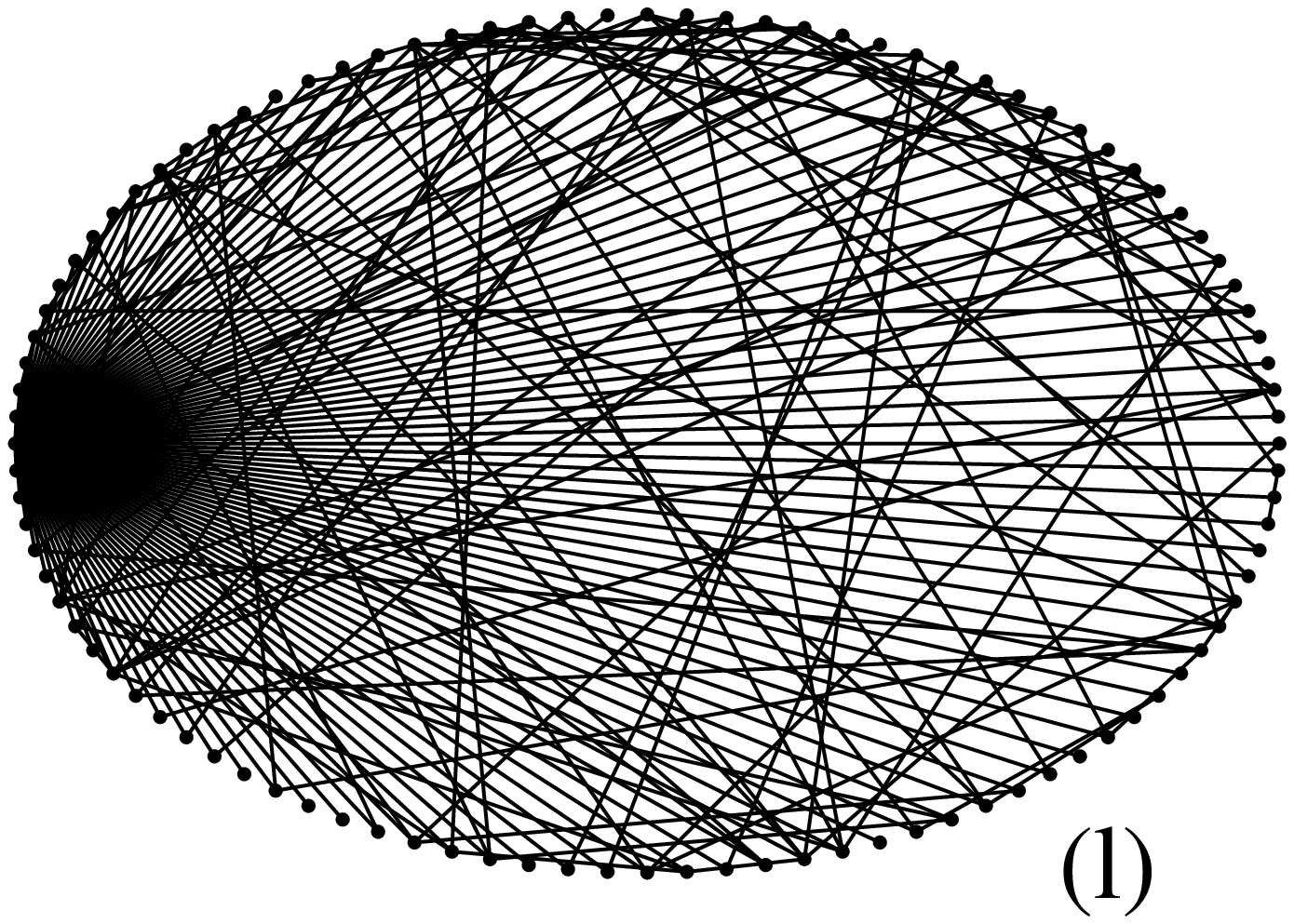,width=4.8cm,height=4.8cm}}  \caption{The ring lattice displays illustrate the evolution of hubs as 
$\lambda$ is varied  over the [0,1] range for an $n=100, k=4$ optimized
network. Very short inter-hub links cannot be distinguished apart from 
local vertex connectivity, however longer range inter-hub links are 
clearly visible. Distinct hub centres illustrate the presence of hubs, as
well as their variation in size and number. The single hub centre at 
the universal hub limit is clearly illustrated. $\lambda$:
(a) 0.0, (b) $5 \times 10^{-4}$, (c) $5 \times 10^{-3}$, 
(d) $1.25 \times 10^{-2}$, (e) $2.5 \times 10^{-2}$, (f) $5 \times 10^{-2}$, 
(g) $1.25 \times 10^{-1}$, (h) $2.5 \times 10^{-1}$, (i) $5 \times 10^{-1}$,
(j) $7.5 \times 10^{-1}$, (k) $8.5 \times 10^{-1}$, (l) 1.0.} 
\label{fig:hub variation}
\end{figure}

\begin{figure}[!htbp]
\centerline{\psfig{figure=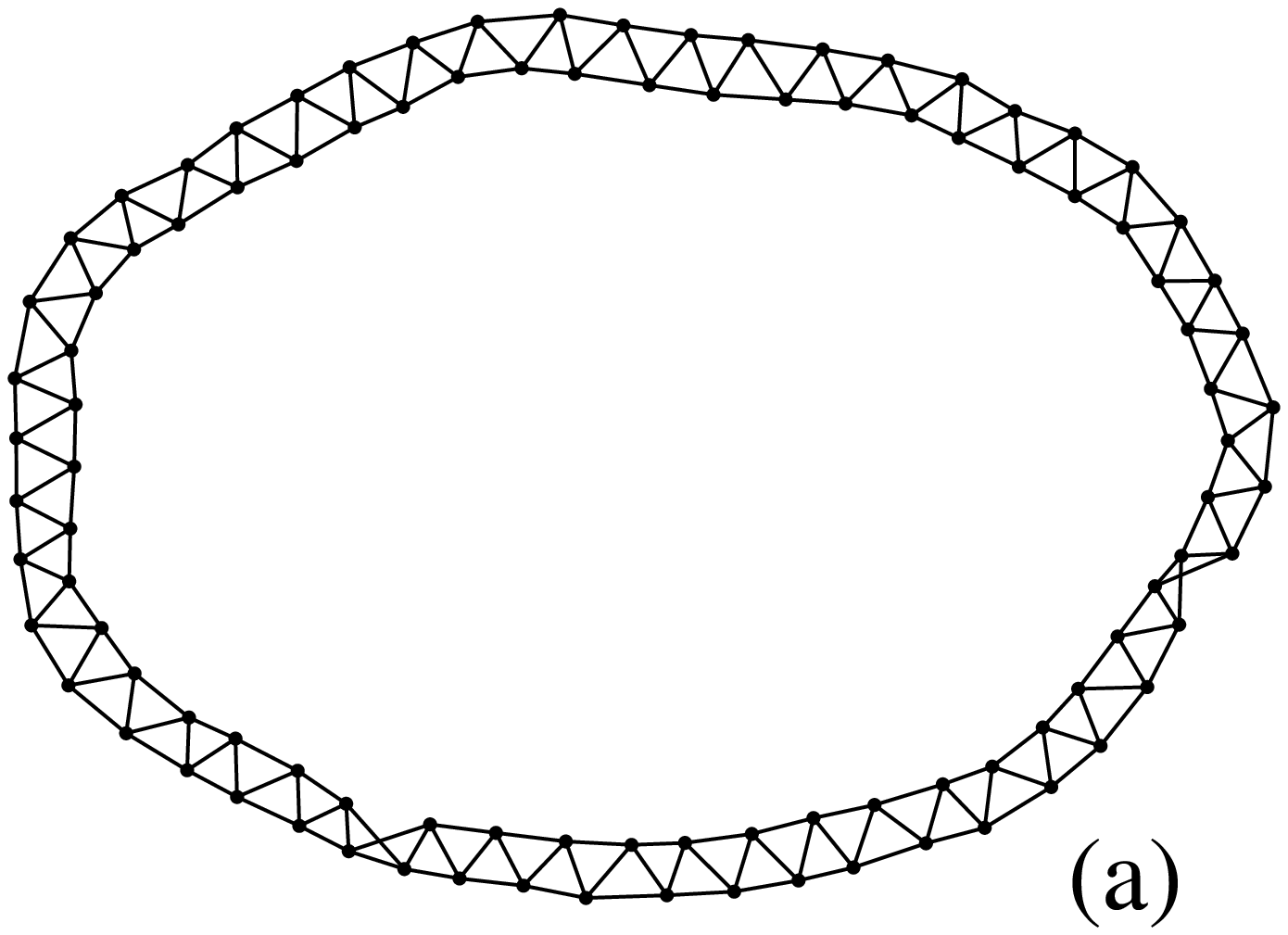,width=4.8cm,height=4.8cm}
            \psfig{figure=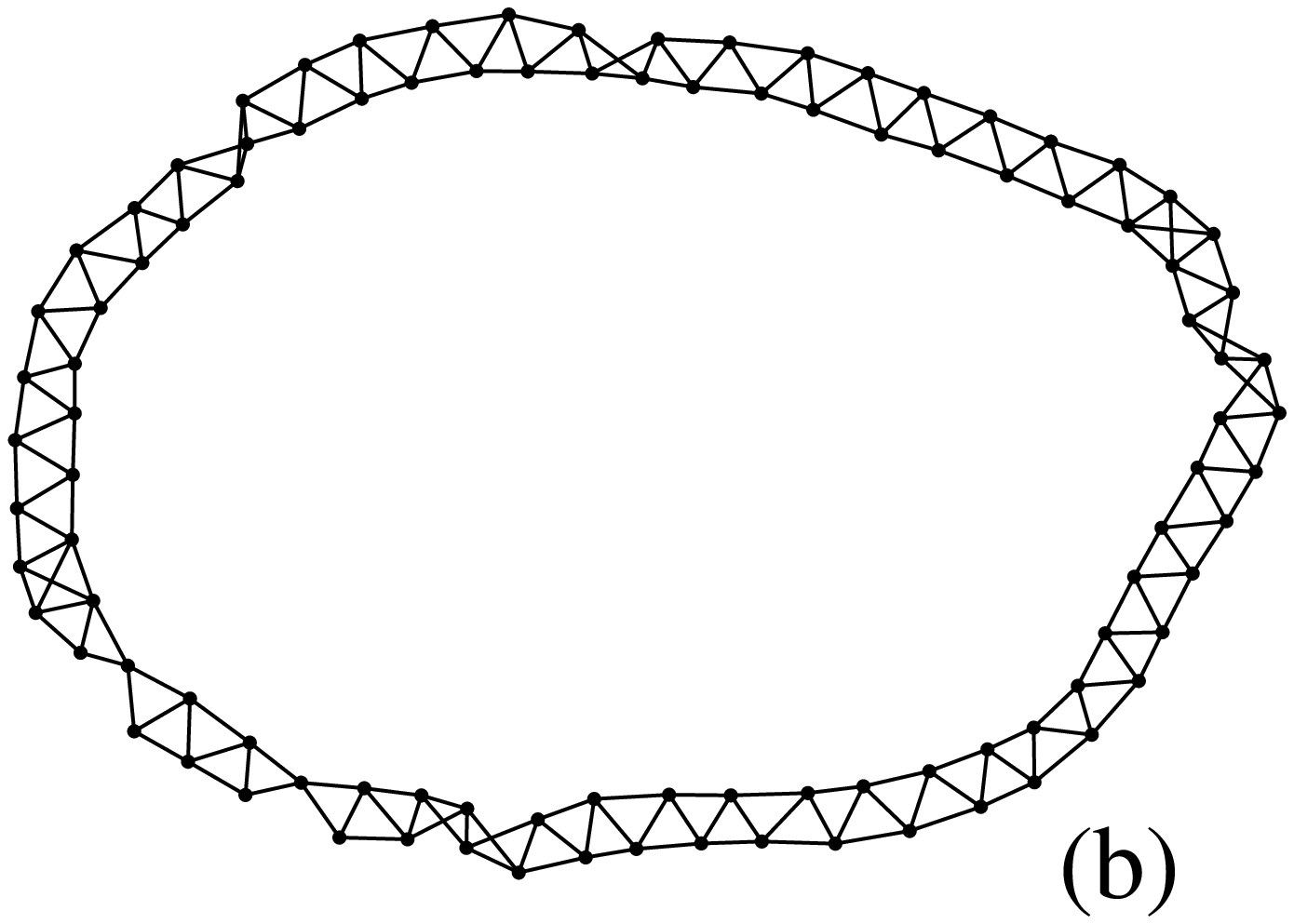,width=4.8cm,height=4.8cm}
            \psfig{figure=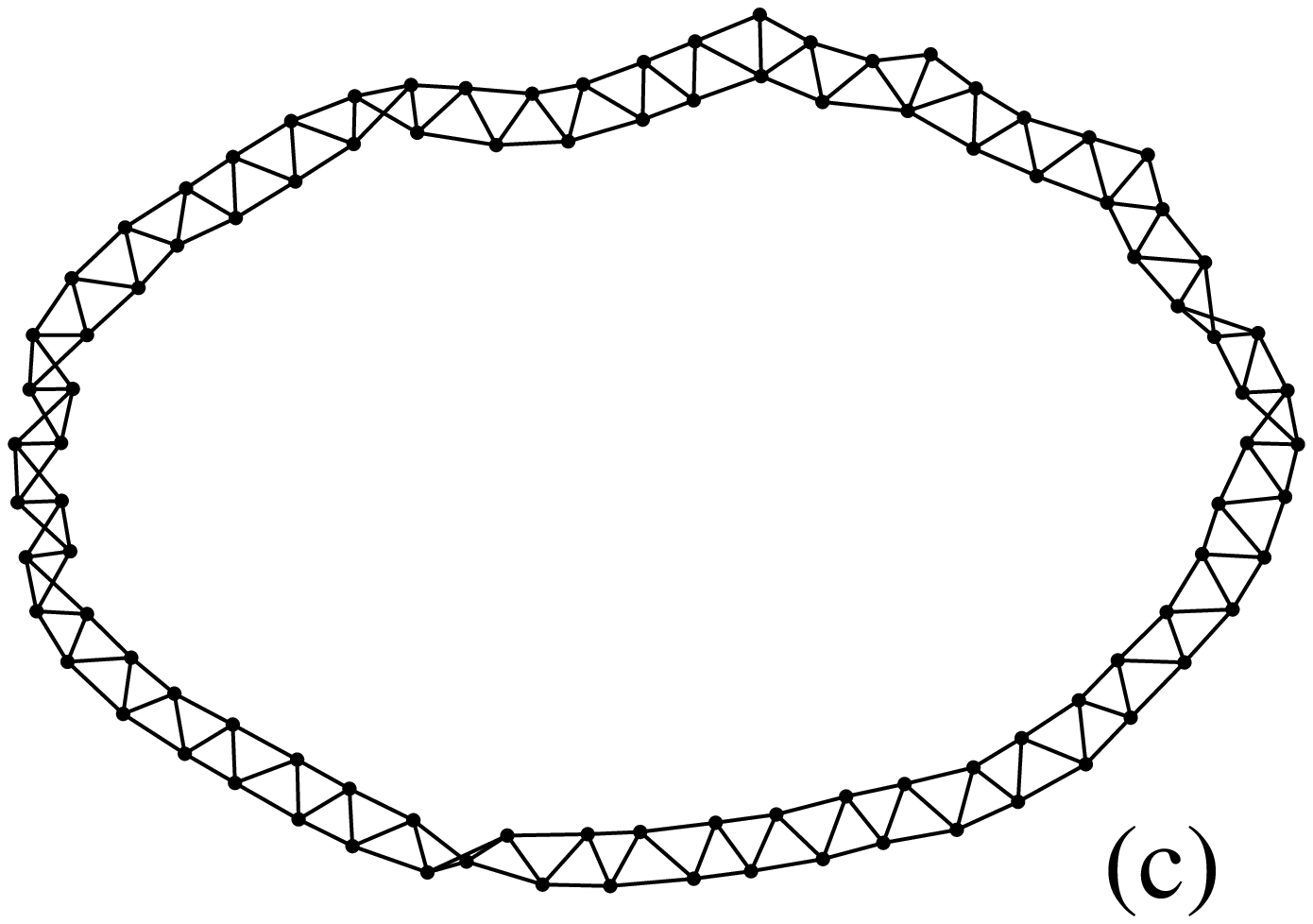,width=4.8cm,height=4.8cm}}
\centerline{\psfig{figure=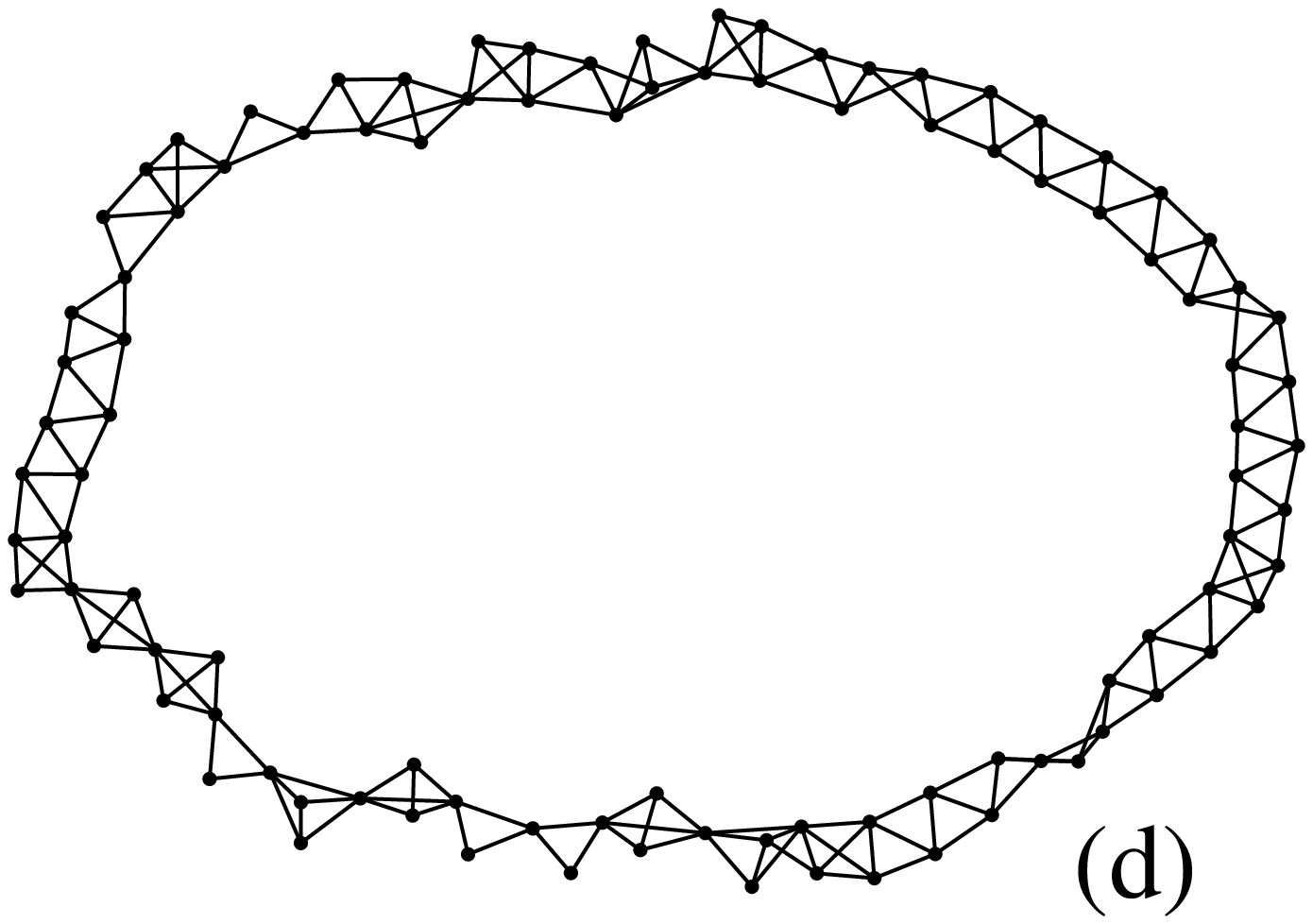,width=4.8cm,height=4.8cm}
            \psfig{figure=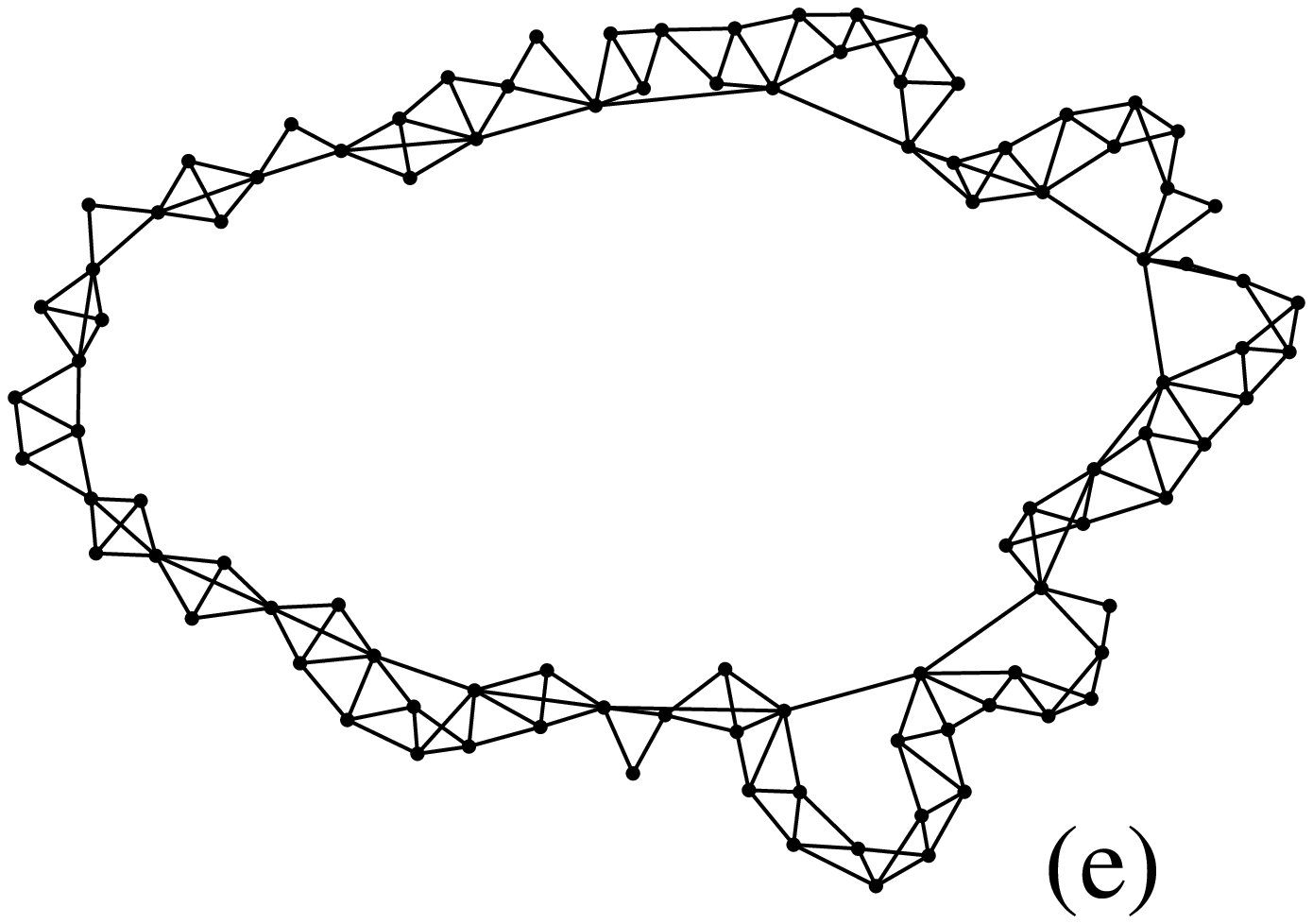,width=4.8cm,height=4.8cm}
            \psfig{figure=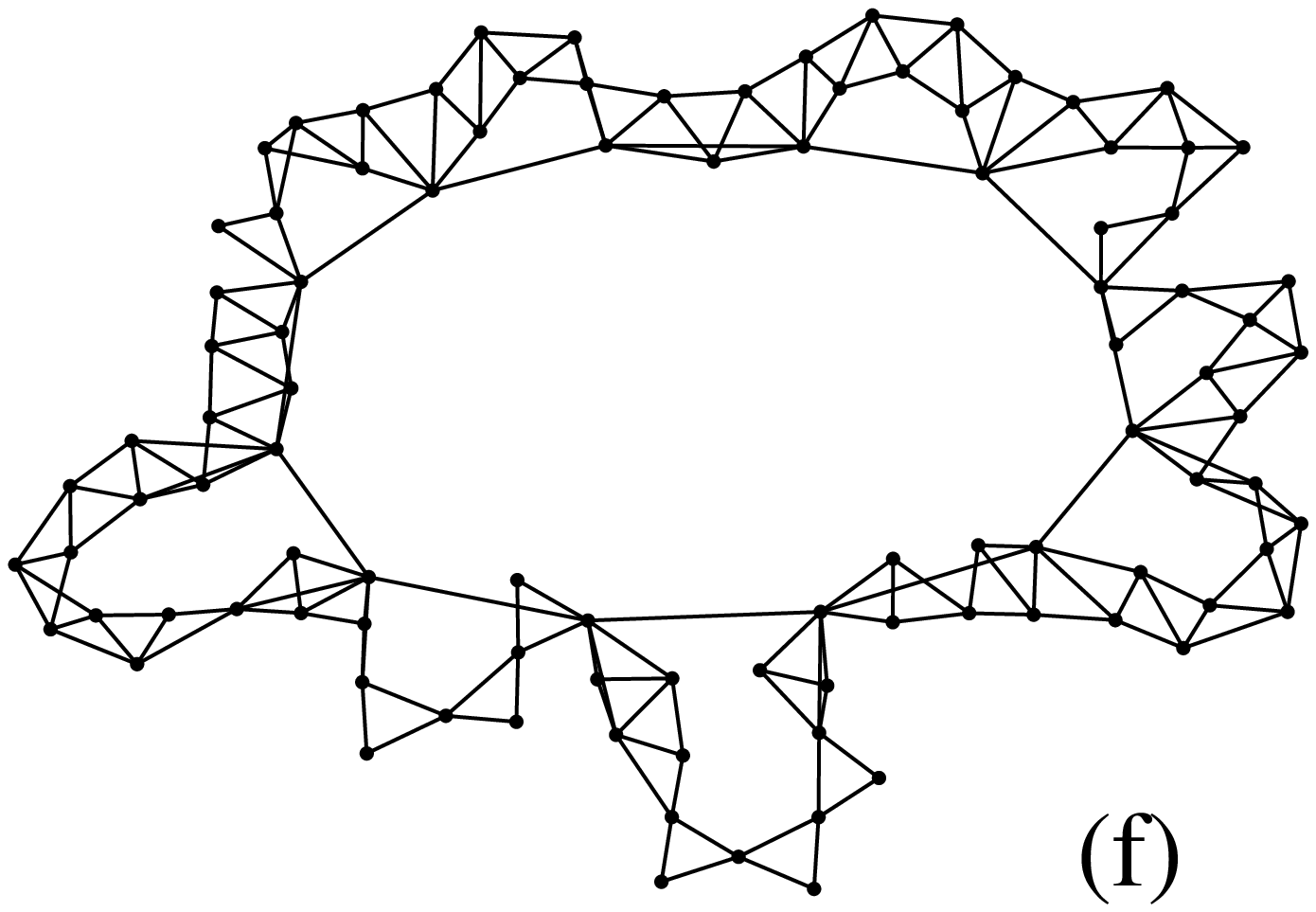,width=4.8cm,height=4.8cm}}          
\centerline{\psfig{figure=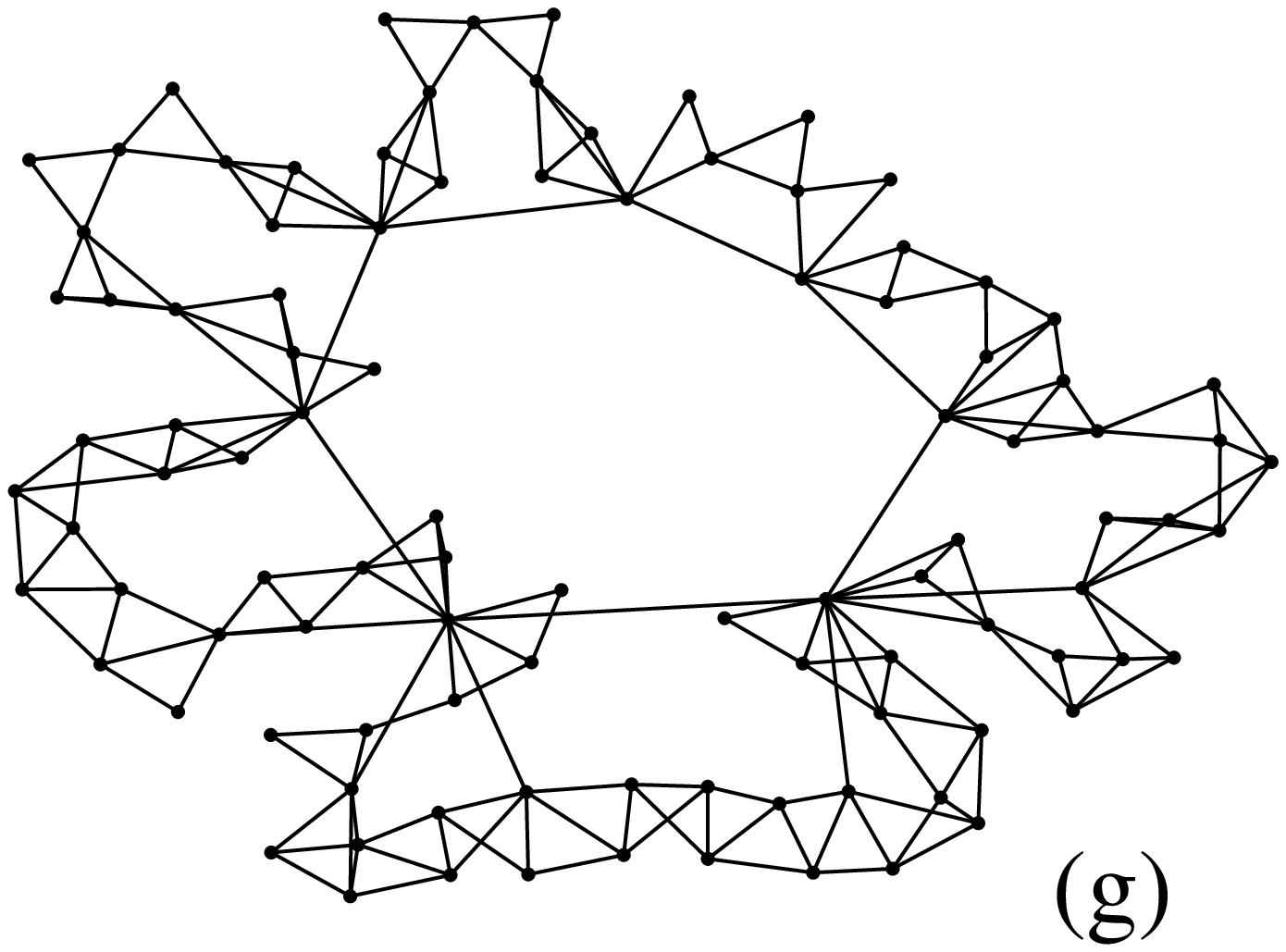,width=4.8cm,height=4.8cm}
            \psfig{figure=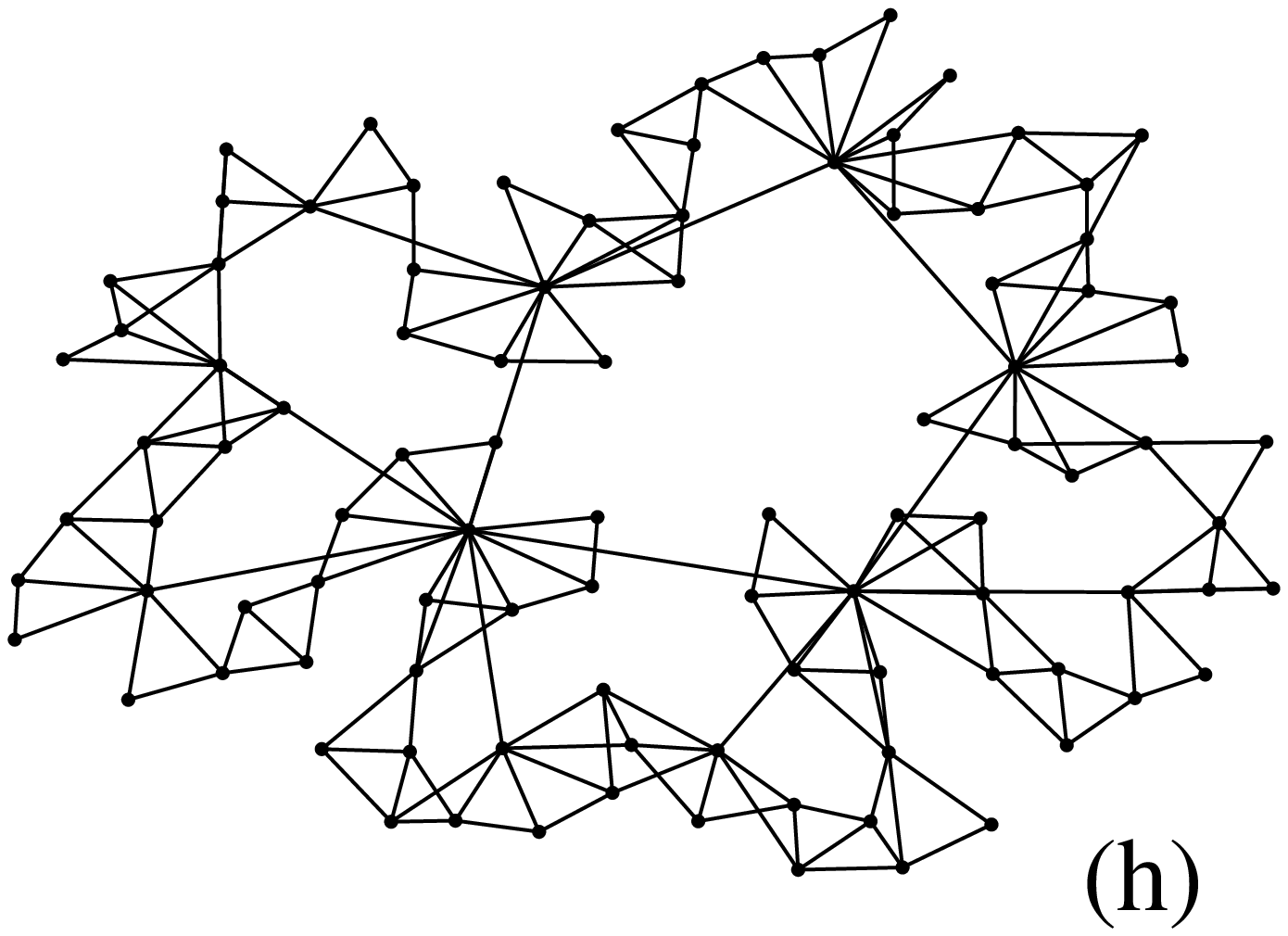,width=5.5cm,height=4.8cm}
            \psfig{figure=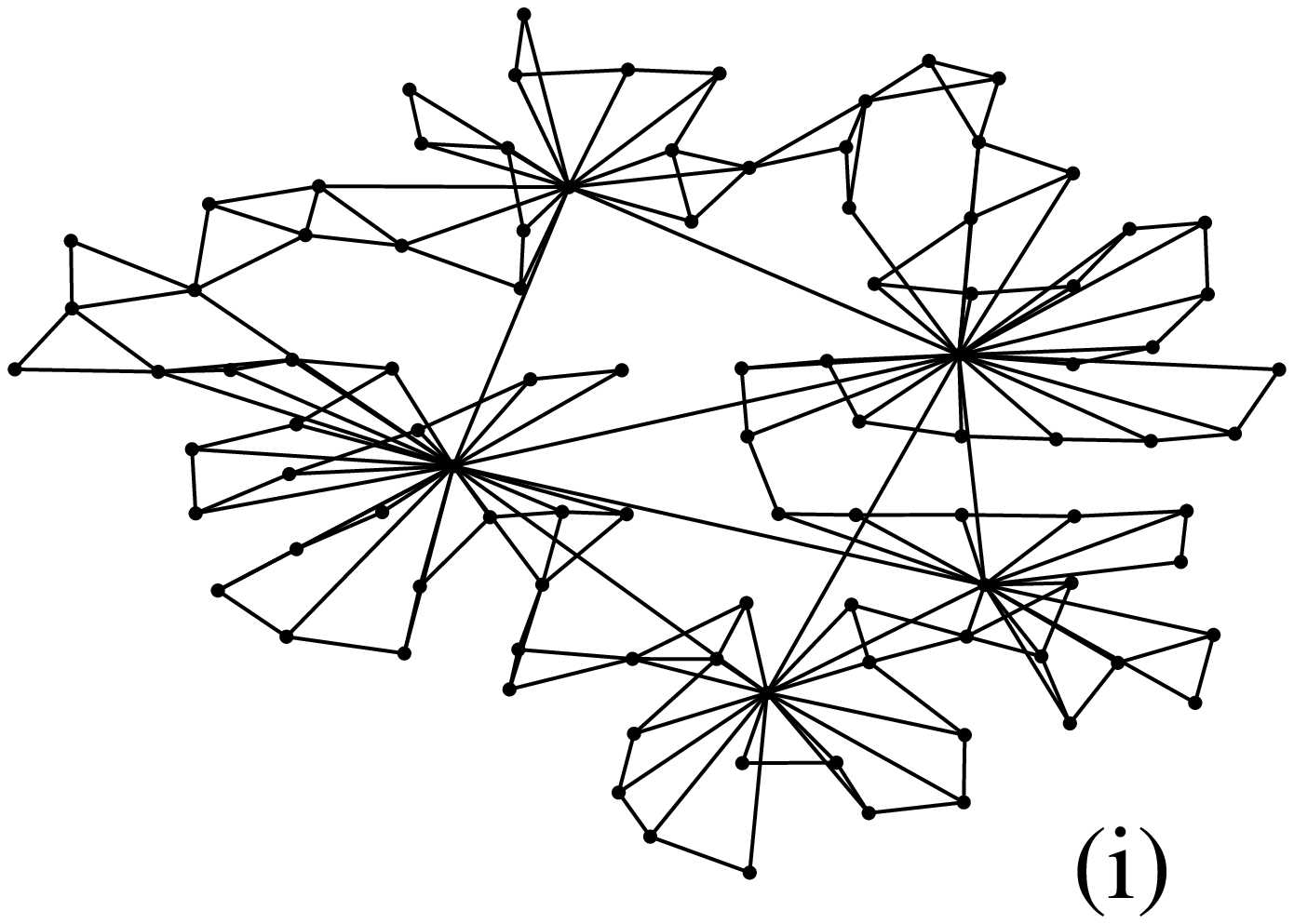,width=4.8cm,height=4.8cm}}
\centerline{\psfig{figure=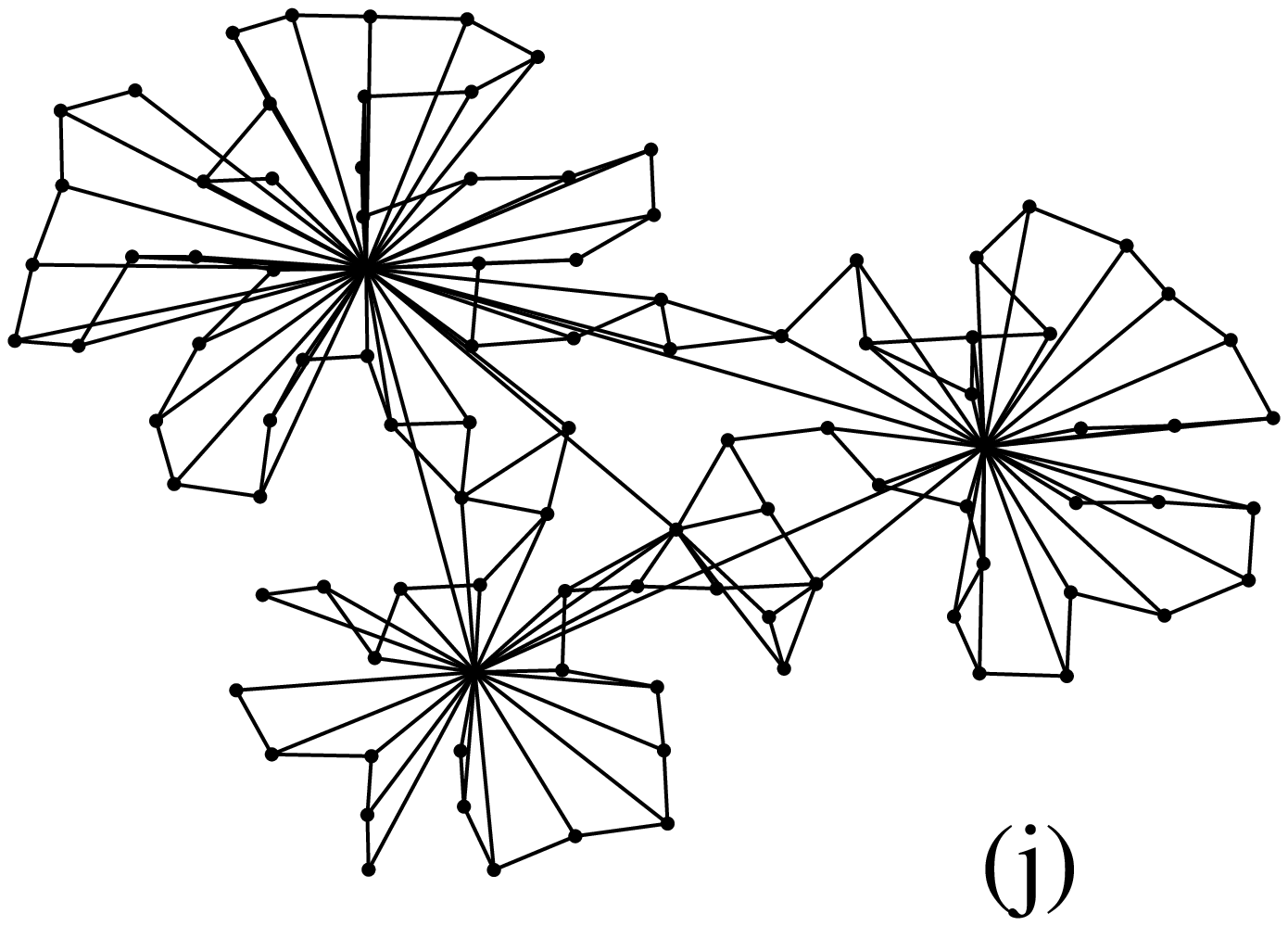,width=4.8cm,height=4.8cm}
            \psfig{figure=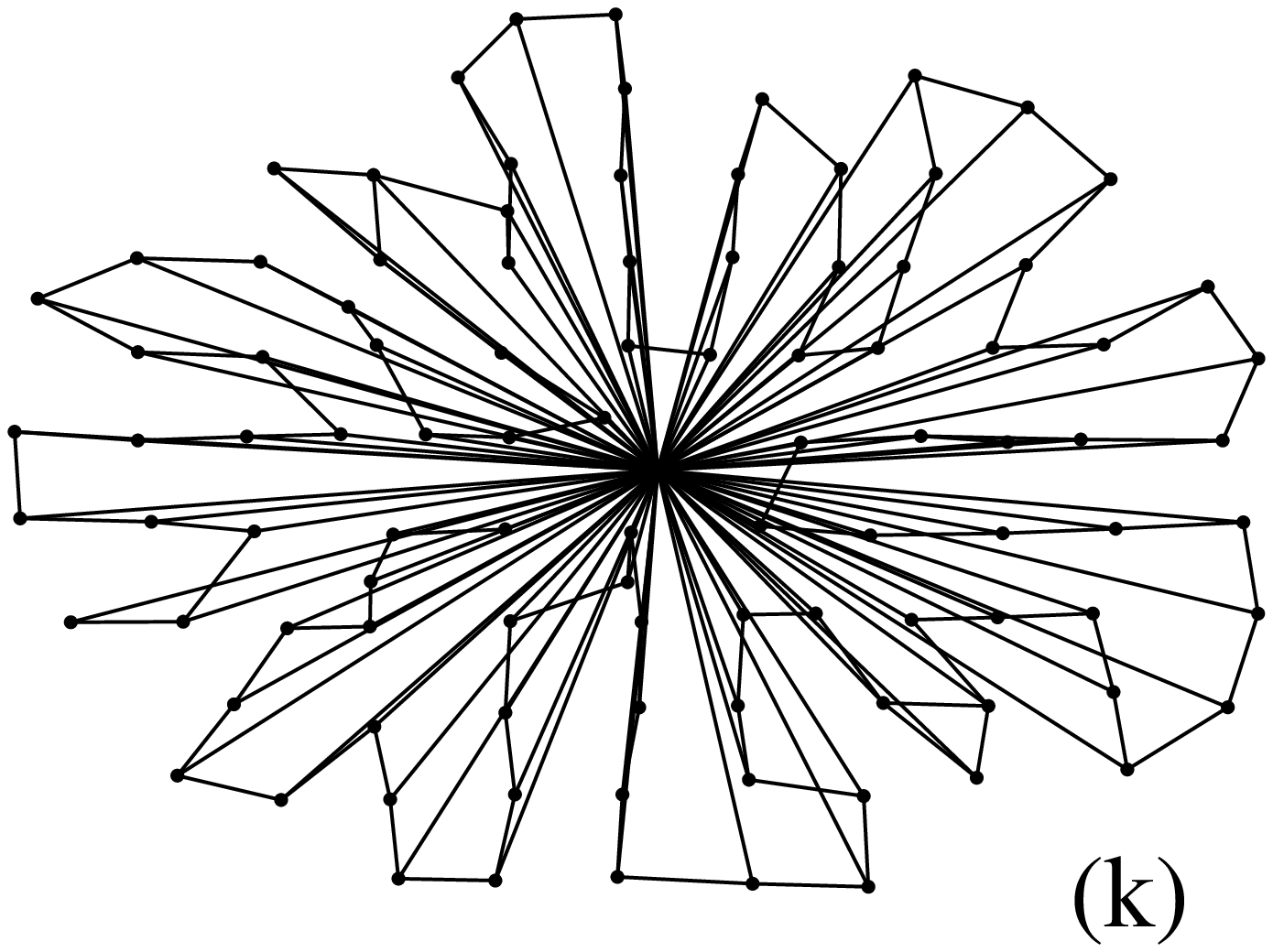,width=4.8cm,height=4.8cm}
            \psfig{figure=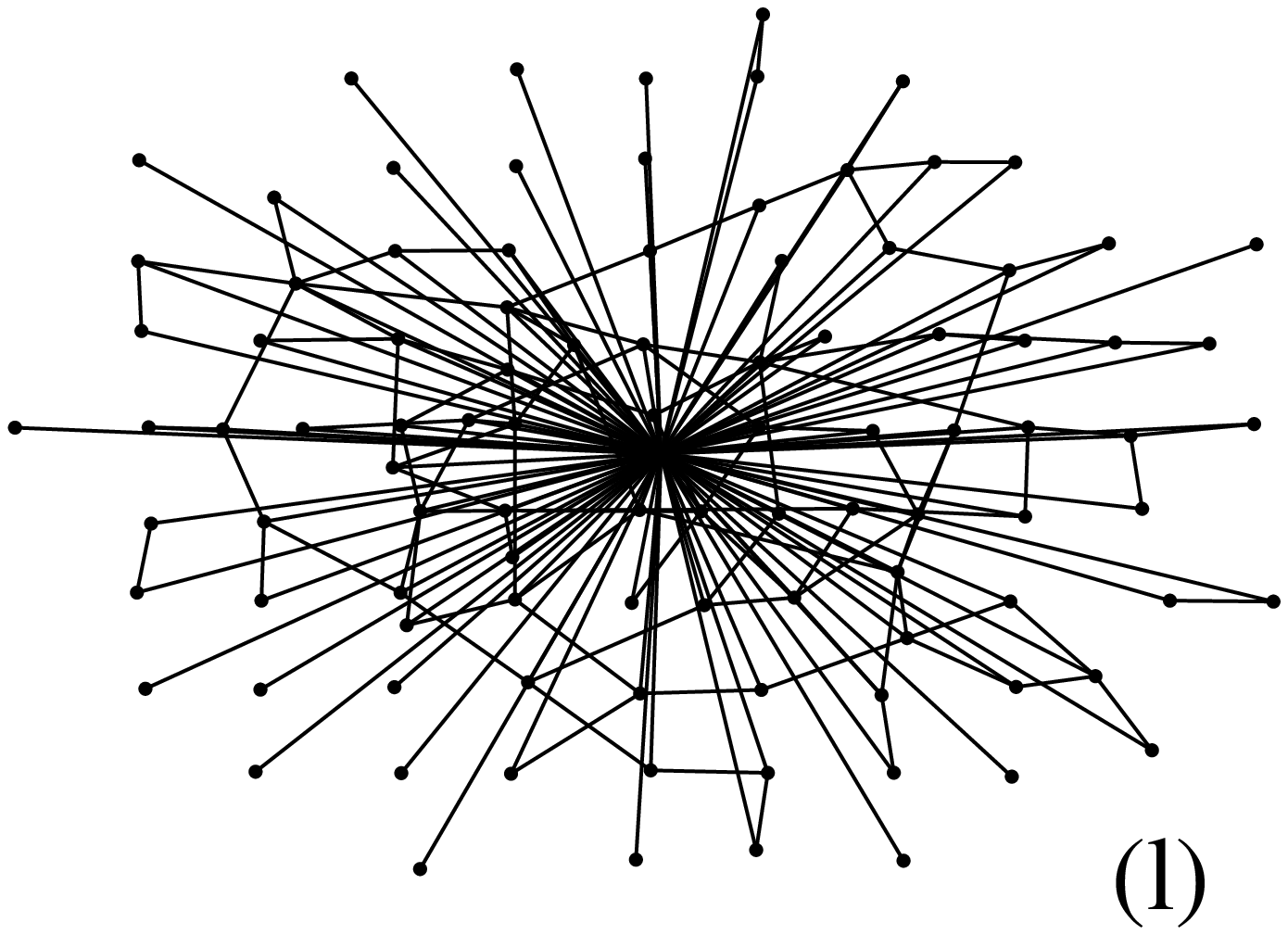,width=4.8cm,height=4.8cm}}
\caption{Illustrates the evolution of hubs for an $n=100, k=4$ optimized 
network as $\lambda$ is varied over the same [0,1] range as the previous
figure. The same networks are displayed as 2d-graphs using a graph generator 
with a spring embedder. Now since vertices are not displayed as
being fixed along a ring lattice, vertex interconnectivity, as well as the
emergence of hubs and their variation in size and number is well illustrated. 
$\lambda$:
(a) 0.0, (b) $5 \times 10^{-4}$, (c) $5 \times 10^{-3}$, 
(d) $1.25 \times 10^{-2}$, (e) $2.5 \times 10^{-2}$, (f) $5 \times 10^{-2}$, 
(g) $1.25 \times 10^{-1}$, (h) $2.5 \times 10^{-1}$, (i) $5 \times 10^{-1}$,
(j) $7.5 \times 10^{-1}$, (k) $8.5 \times 10^{-1}$, (l) 1.0.}
\label{fig:2d hub variation}
\end{figure}

The formation of the universal hub at sufficiently large $\lambda$ is
not surprising, since it can be shown that for a network that
minimizes $L$ and employs only rewirings, a universal hub will effect
the largest minimization. The formation of {\em multiple} hubs
however is due to the role played by $W$ in the optimization, which is
to constrain the physical length of edges, and therefore, the size of
hubs. As the hubs grow, whenever the cost of edges from the hub
centre to farthest nodes become high, the edges break away resulting
in multiple hubs. Thus, high wiring cost prevents the formation of
very large hubs, and controls both the size and number of hubs.
Figures~\ref{fig:hub variation} and \ref{fig:2d hub variation},
demonstrate the evolution of hubs in an $n=100, k=4$ optimized network
as $\lambda$ is varied between 0 and 1. While Fig.\ \ref{fig:hub variation}
uses ring-lattice displays to illustrate the evolution, 
Fig.\ \ref{fig:2d hub variation} illustrates the same
networks as 2d-displays. In the ring-lattice displays, vertices are fixed
symmetrically around the lattice, with hub centres and long-range inter-hub
links being clearly visible. The 2d-displays are generated by a graph drawer 
which uses a spring embedder to clearly demonstrate vertex interconnectivity.
Now, with vertices no longer fixed along a ring-lattice, short-range inter-hub
links can be distinguished apart from local connectivity. 

\subsection{Hub evolution}

We now detail the evolution of hubs using the edge scale distribution
shown in Fig.\ \ref{fig:optEdgeScaleDist}, and hub variation described 
by Figs.\ \ref{fig:hub variation} and \ref{fig:2d hub variation}. 
All three figures show the same $n=100$ and $k=4$ network at various 
$\lambda$.

In Figs.\ \ref{fig:optEdgeScaleDist} and \ref{fig:2d hub variation}(a),
the optimization results in a near regular network, with hardly any
hubs. When the cost reduces slightly to allow for an increase in edge
wiring, small hubs are formed. For increasing, but very small
$\lambda$, Figs.\ \ref{fig:optEdgeScaleDist}(b-c) show the edges
to be almost entirely concentrated in the unit length scale, with very
few longer edges. The non-unit length scale edges account for very few and
very small hubs, as illustrated in Figs.\ \ref{fig:2d hub variation}(b-c). 
Due to their small size, and very short inter-hub links,
the hubs are indistinguishable from local vertex connectivity in the
ring lattice displays in Figs.\ \ref{fig:hub variation}(a-d).

A slight fall in the wiring cost permits an increased number of hubs.
The high cost of wiring constrains hubs to be still rather small. 
Hence, the distribution of scales in
Fig.\ \ref{fig:optEdgeScaleDist}(d) still shows only two length
scales. However, there is a marked increase in edges of the second
length scale. The effort towards minimizing $L$, ensures that the few
hubs are bunched close together so that short inter-hub links can be
used to enable the maximum distance contraction possible 
(Fig.\ \ref{fig:2d hub variation}(d)).

When further reduction in cost permits increased wiring, it is mostly
the inter-hub links that take advantage of the reduced cost to enable
hubs to be scattered over the entire network.
Figure~\ref{fig:optEdgeScaleDist}(e) shows clearly the multiple
length scales generated by inter-hub links. The marked increase in
the range of the inter-hub links (Fig.\ \ref{fig:hub variation}(e)),
allows them for the first time to be visible in the ring-lattice
plots. Figure~\ref{fig:2d hub variation}(e) shows that there is not
much variation in the hub size, except for the longer range of the
inter-hub links.

Figures~\ref{fig:optEdgeScaleDist} and \ref{fig:hub variation}(f-h)
demonstrate that as $\lambda$ increases further, the length and number
of far edges are progressively less constrained, and the extended length 
permits larger and many more hubs. Vertices lose their local nearest-neighbour
interconnectivity as hubs centres dominate in connectivity 
(Figs.\ \ref{fig:2d hub variation}(f-h)). However, as the size of hubs 
increases, they are consequently reduced in number. In 
Figs.\ \ref{fig:2d hub variation}(i-k) one observes efforts towards 
a uniform reduced local connectivity. The number of inter-hub
links increases to yield greater inter-hub distance contraction.

Figures~\ref{fig:2d hub variation}(i-j) are marked by a sharp reduction in the 
number of hubs as the hubs balloon in size. This evolution culminates in the
emergence of the universal hub, (Fig.\ \ref{fig:2d hub variation}(k)), a 
single hub of connectivity. 
The formation of 
edges between the hub centre, and
all the other $n-1$ vertices, as illustrated in Fig.\ \ref{fig:hub
variation}(k), results in a uniform distribution of non-unit length
scale edges. Wiring, which is still associated with a cost, albeit
small, ensures that the remainder of the edges are entirely local, as
can be observed from the distribution in Fig.\ \ref{fig:optEdgeScaleDist}(k).

Figure~\ref{fig:hub variation}(l) demonstrates that when $\lambda = 1$,
the universal hub is retained. However, due to the absence of any effort
towards minimal wiring, edges are uniformly distributed across the entire
length scale range as shown in Fig.\ \ref{fig:optEdgeScaleDist}(l).
The loss in local connectivity can be clearly seen in comparison to
Figs.\ \ref{fig:hub variation} and \ref{fig:2d hub variation}(k). 
Optimization towards minimizing 
only $L$, results in the {\em re-emergence} of multiple hubs, but the
removal of the constraint on wiring allows hubs to be composed of 
largely non-adjacent vertices.

In conclusion, during the evolution of hubs illustrated in figures
(a-l), as the cost of wiring is decreased, the following sequence is seen :
\begin{my_item} 
\item Hubs emerge, and grow in size and number
\item Increase in the range and number of inter-hub links
\item Subsequent reduction in the number of hubs 
\item Formation of a universal hub 
\item Hubs re-emerge accompanied by a loss in local vertex interconnectivity,
while the universal hub remains 
\end{my_item}

\section{Optimization and the WS model: Some comparisons}

For the remaining part of this section, we present further results,
but against the backdrop of the WS model. To define small-world
behaviour, two ingredients were used by Watts and Strogatz. The first
was the characteristic path length, a global property of the graph,
while the second, the clustering coefficient, $C$, is a local property
which quantifies neighbourhood  `cliquishness'. Associated with each 
vertex $v$, is its neighbourhood, $\Gamma_v$, the $k_v$ vertices to which it 
is directly connected, and among which there can be a maximum of 
$k_v(k_v-1)/2$ connections. $C_v$, the clustering
coefficient of $v$, denotes the fraction of the links actually present 
among its neighbours, defined as
\begin{eqnarray*}
C_v = \frac{|E(\Gamma_v)|}{\scriptsize \left( \begin{array}{c} k_v \\
2 \end{array} \right)}, \end{eqnarray*} 
while $C$ is $C_v$ averaged over all $v$.

The WS and optimization models are compared with respect to their 
normalized small-world characteristics. 
In addition, we study their 
different behaviours with respect to normalized wiring and degree. All 
results are obtained using an $n=100, k=4$ network. 
Each plot is the result of averaging over 40 simulation runs.

\subsection{Characteristic path length}

\begin{figure}[!htbp]
\centerline{\psfig{figure=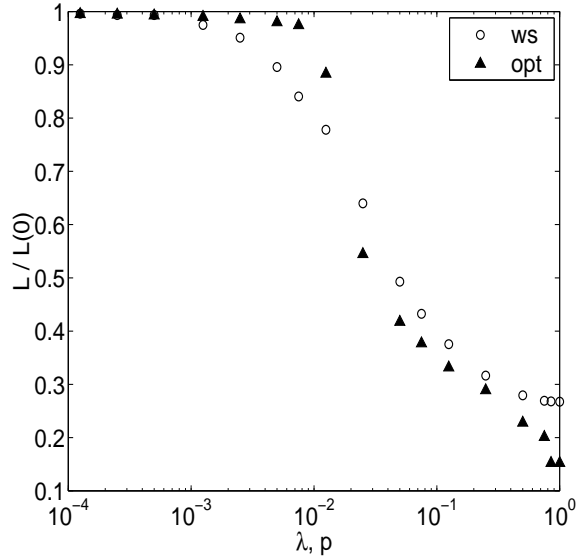,width=7.6cm,height=7.5cm}}
\caption{Variation in the normalized characteristic path length, $L/L(0)$,
versus $p$ and $\lambda$, for the WS model and optimization model
respectively.}
\label{fig:wsOptL}
\end{figure}

We begin our comparison with the characteristic path length, the
parameter whose smallness gives these networks their name.
Figure~\ref{fig:wsOptL} compares $L$ for the WS and optimized
models. The control parameters in the two models, $\lambda$ the
optimization parameter, and $p$ the WS parameter, are similar in that
they both control the introduction of far edges. It should be
remembered though, that while $p$ controls only the {\em number} of
far edges, allowing their length scales to be uniformly distributed
across the entire range, $\lambda$ constrains not only the number,
but also the physical {\em length} of far edges.

In both cases, $L$ shows a sharp drop that signifies the onset of
small-world behaviour. However, in contrast to the gradual drop effected by 
the random assortment of rewired edges in the WS model, 
the drop due to hub formation is much sharper.
Although its initial reduction is 
smaller due to the additional constraint on edge length, 
its final value  is much lower than the WS random graph limit.

The variation in $L$ resulting from optimization, can be understood 
from the role played by the hub centres in contracting distance between pairs of
vertices.  
Before the cliff, the hubs being few and very small, effect a very slight
distance contraction. The tip of the cliff forms due to a marked increase in
hubs, while the sharp drop occurs when extended range inter-hub links yield
a pronounced distance contraction between many distant hubs and 
their widely separated neighbourhoods.
The transition from many, small hubs to much larger and consequently
fewer hubs, results in the gradual reduction in $L$. Finally, on the 
emergence of the universal hub, which has no counterpart in the WS
model, the single hub centre contracts the
distance between {\em every} pair of vertices, resulting in an average 
distance less than 2. 

\subsection{Clustering coefficient}

\begin{figure}[!htbp]
\centerline{\psfig{figure=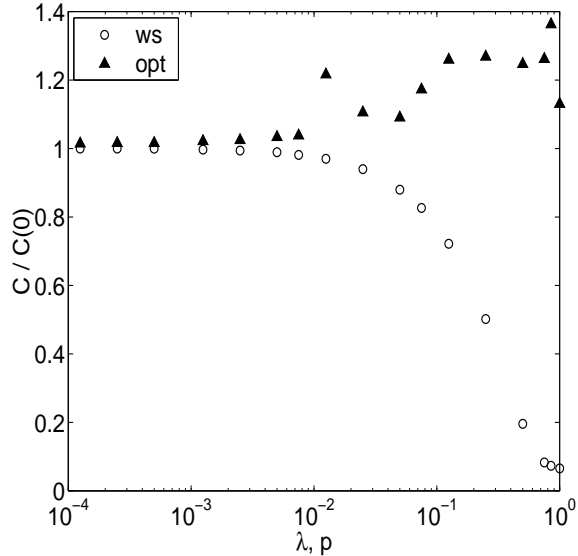,width=7.6cm,height=7.5cm}}
\caption{Comparison between the WS model and optimization model with respect to
the variation in their normalized clustering coefficient, $C/C(0)$.}
\label{fig:wsOptC}
\end{figure}

Figure \ref{fig:wsOptC}, which compares the variation in clustering coefficient 
for the two
models, shows far more interesting behaviour. The drop in local
connectivity that is seen in the WS model does not occur {\em at all}
for the optimized network because of the formation of hubs. Although
the clustering coefficient was not a characteristic that was sought to
be maximized, high cliquishness emerges. Figure~\ref{fig:wsOptC}
shows that the formation of hubs sustains the clustering coefficient
at a value higher than that for the corresponding regular graph,
unlike the WS model. Thus, the similarity between $p$ and $\lambda$
as control parameters is only valid for $L$.

Before a more detailed analysis of Fig.\ \ref{fig:wsOptC}, we discuss the
clustering
coefficient further. For a vertex $v$, its neighbourhood size $k_v$,
plays a significant role. The smaller is $k_v$, 
the smaller the number of possible intra-neighbourhood edges. Hence,
vertices which lose in connectivity, gain in
cliquishness. In a similar manner, vertices which gain in
connectivity, lose in cliquishness because of their larger
neighbourhood size. This is because, although the vertices have a
larger number of intra-neighbourhood edges, they form a smaller
fraction of the total number of possible edges. At the universal hub
limit, the hub centre has the least clustered neighbourhood owing to
the fact that all the remaining $n-1$ vertices form its neighbourhood.
The clustering coefficient can be shown to be approximately
$(k-2)/n$. Although the average degree remains unchanged, the varying hub size
and number can influence which neighbourhoods dominate the average clustering
coefficient.

In addition, a factor which influences the clustering within
neighbourhoods, is the inclusion of a hub centre to a neighbourhood.
The effect on clustering differs depending on the range of the link
between the vertex and the hub centre. If the range of the link is
large and the vertex lies outside the hub, then a far away vertex is
being included into an otherwise locally connected neighbourhood. The
hub centre has little or no association with the remaining neighbours
and so it lowers the average cliquishness. However, if the vertex
lies within the hub, it amounts to including a node which is connected
to all, or a large fraction of its neighbours. Hence, its
neighbourhood becomes more clustered. This effect is more pronounced
when (1) the neighbourhood size is small, (2) the vertex includes more than one
hub centre in its neighbourhood, and (3) the hub whose centre is being included
is composed of largely local neighbours.
Thus, unlike $L$ which is tuned by a single parameter, $C$ is
controlled by many more, a point that we will return to shortly in the
analysis of Fig.\ \ref{fig:wsOptC}.

From Fig.\ \ref{fig:wsOptC} we see as expected for the WS model, that 
$C(p)$ falls as
approximately $C(0)(1-p)^3$ \cite{3:barrat}, and eventually drops
almost to zero for the completely randomized graph. This is due to
the increasing number of inclusions of random far nodes, into
otherwise locally connected neighbourhoods. In contrast, the presence
of hubs in the optimization model ensures that $C(\lambda)$ never falls below
$C(0)$; reaching its maximum when the network converges to a universal
hub.

Keeping in mind the evolution of the optimized network shown in 
Figs.\ \ref{fig:hub variation} and \ref{fig:2d hub variation}
we can understand qualitatively the behaviour of $C(\lambda)$ in 
Fig.\ \ref{fig:wsOptC}. When $\lambda$ is small, the network is
dominated by regular neighbourhoods (Fig.\ (a-c)). As described
earlier, vertices adjacent to hub centres gain in cliquishness due to their
reduced neighbourhood sizes. Despite there being just a few small hubs, since
there are more reduced connectivity vertices than hub centres, the average $C$
is raised slightly above that of a regular graph.

With a slight increase in $\lambda$, 
Figs.\ \ref{fig:hub variation} and \ref{fig:2d hub variation}(d) shows a sharp increase in $C$. At
this point, the marked increase in hubs, with only a slight increase
in size, results in a pronounced increase in the number of reduced
connectivity vertices. Many of these vertices have neighbourhoods which are 
completely clustered (called cliques), since in addition to their reduced 
size, they include one or more hub centres into their neighbourhood.
Cliques, not surprisingly, dominate the average resulting in the large
jump in $C$. However, the emergence of long range inter-hub links in 
Figs.\ \ref{fig:hub variation} and \ref{fig:2d hub variation}(e-f) results in lowering $C$. Their introduction causes: (1) hub 
centres to
have lowered cliquishness owing to the inclusion of distant nodes into their
neighbourhoods, and (2) some reduced connectivity neighbourhoods to no 
longer be complete cliques due to the inclusion of the centre of a hub, which
has lost local neighbours to inter-hub neighbours.

Across Figs.\ \ref{fig:hub variation} and \ref{fig:2d hub variation}(g-h), $C$ rises again due to increased cliques generated
not only 
by the increased number and size of hubs, but also because their larger
size allows for the inclusion of local neighbours once again.
However, in Figs.\ \ref{fig:hub variation} and \ref{fig:2d hub variation}(i-k), while the few hubs gain in connectivity (and
consequently lose in cliquishness),
the remaining vertices veer towards uniformity in reduced connectivity. 
The resulting marked reduction in the number of cliques, accounts for the
slight drop at Figs.\ \ref{fig:hub variation} and \ref{fig:2d hub variation}(i), while the near uniform reduced connectivity serves to
further raise $C$.  
Finally, in the universal hub limit, (Figs.\ \ref{fig:hub variation} and \ref{fig:2d hub variation}(k)), 
all vertices have a uniform reduced vertex connectivity, at the expense 
of the single hub centre. Having gained in cumulative connectivity, the 
hub centre has a very low clustering coefficient of approximately $(k-2)/n$. 
However, the remaining reduced sized neighbourhoods and their inclusion of the 
hub centre, ensures the average $C$ shoots up to its maximum.   

In Figs.\ \ref{fig:hub variation} and \ref{fig:2d hub variation}(l), the average clustering falls due to the non-uniformity in vertex
connectivity. However the inclusion of the universal hub centre into every
neighbourhood ensures $C$ does not drop too much. 
The multiple hubs result in a variation in vertex 
connectivity, with hub centres gaining in connectivity at the expense
of others. This leaves a few vertices having a {\em single} connection. 
With a neighbourhood of only 1, and no intra-neighbourhood connectivity,
these vertices are totally unclustered\footnote{Such vertices have $k_v=1$, 
and $|E(\Gamma_v)|=0$, which results in an invalid definition of $C_v$. 
Their clustering coefficient can be taken to be either 0 or 1. To be
noted, is that rather than 0, a value of 1 would result in the average
clustering coefficient being higher than that at the universal hub.} which 
accounts for the drop in $C$.

Thus, we see an interesting inter-play between neighbourhood size, hub 
centre inclusions, and the number and range of inter-hub links. However, 
the data for the variation of $C$ in the optimized model is noisy,
mainly because $k$ is very small. Constraints in computational resources
have forced us to work with small $n$. Further, to maintain the sparseness 
condition of $n \gg k$, a low $k$ was used, which 
does not really satisfy the WS condition that $k \gg 1$. Due to the small $k$,
even a small loss in connectivity, can cause neighbourhood
cliquishness to rise sharply. Although different factors come into
play during the $C$ variation, the spikes are due to the pronounced
effect of reduced connectivity neighbourhoods, and in particular to
those of cliques. The cliques serve to maintain the entire
$C$ variation higher than would probably result for higher $k$. 
Work is in progress to obtain data using large $k$ networks.

\subsection{Wiring cost and Degree}

\begin{figure}[!htbp]
\centerline{\psfig{figure=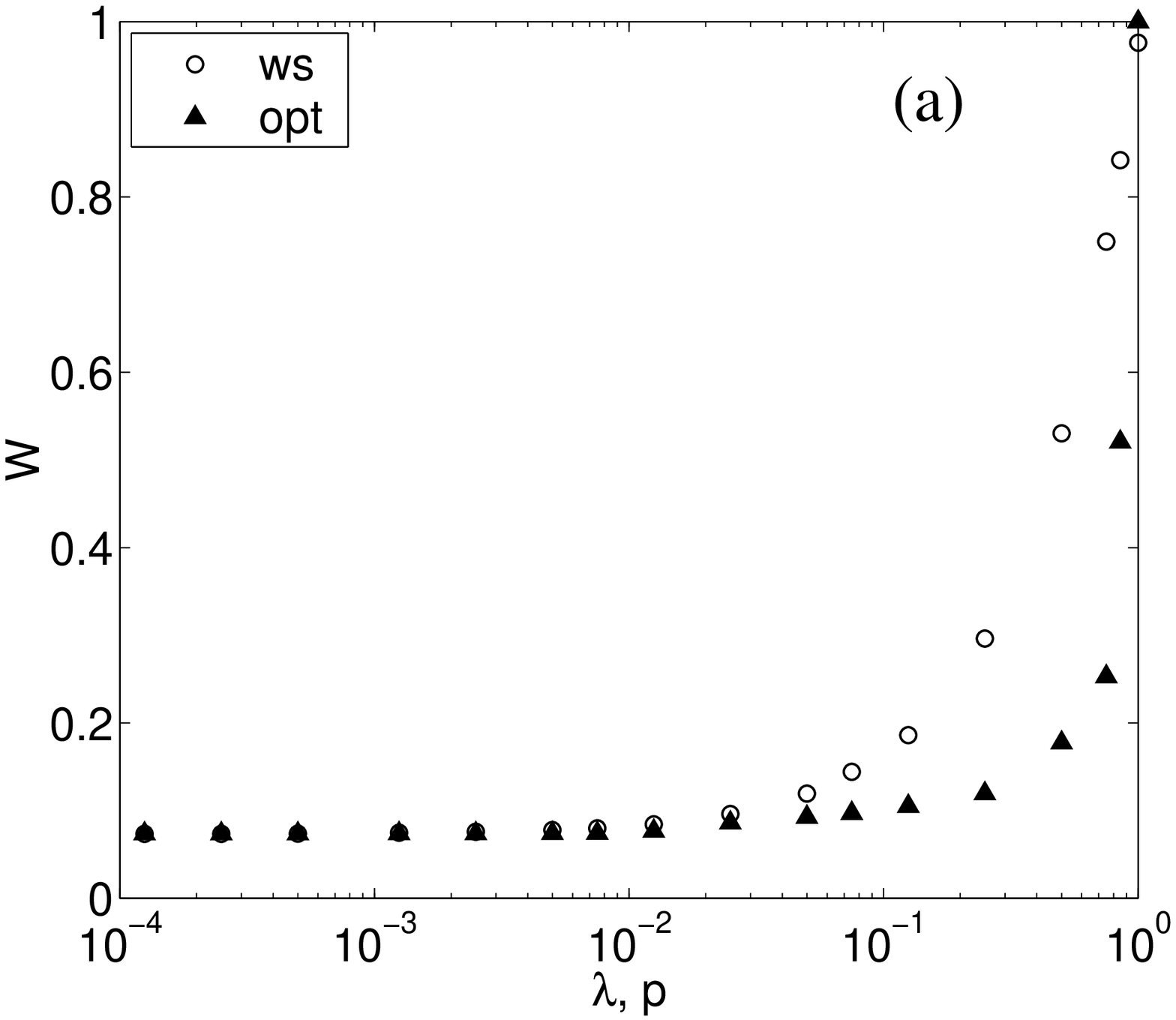,width=7.6cm,height=7.5cm}
            \psfig{figure=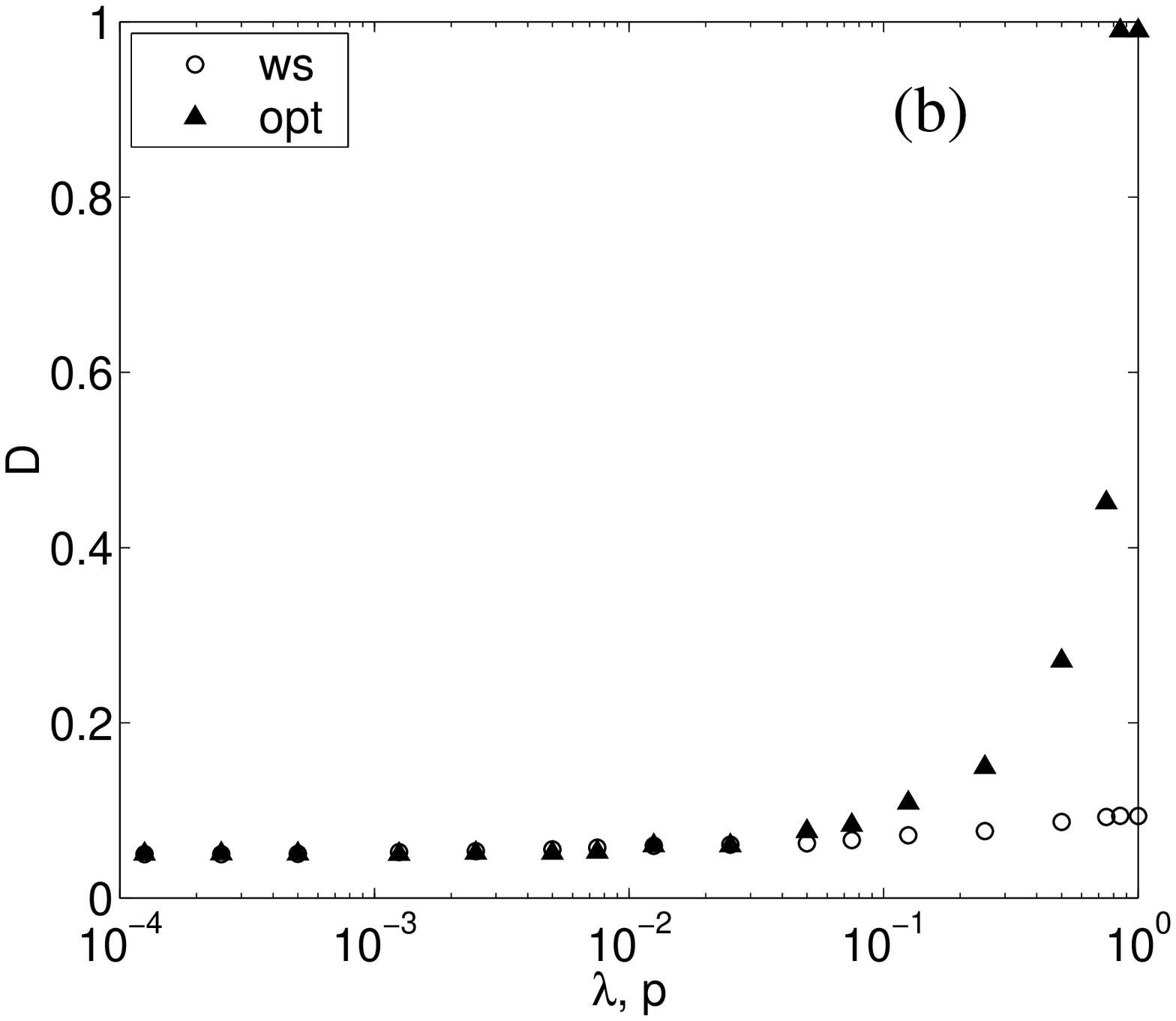,width=7.6cm,height=7.5cm}}
\caption{Comparison between the WS and optimization models versus $p$ and
$\lambda$ respectively. (a) Variation in the wiring cost, $W$, normalized 
by the optimized value of $W(1)$. (b) Variation in the maximum degree, $D$, 
normalized by network size, $n$. $D$ is equivalent to the size of the largest
hub in the optimization model.} 
\label{fig:optWD}
\end{figure}

Figure~\ref{fig:optWD}(a) displays the increase in the cost of wiring, or
alternatively, the amount of wiring, with $p$ and $\lambda$.
The comparison between the optimization model and WS model
illustrates clearly the difference made by the inclusion of the
minimal wiring constraint. For small
$\lambda$, both models exhibit similar wiring cost. At larger
$\lambda$ however, the absence of a similar constraint in the WS model
results in a much greater amount of wiring. The clear advantage exhibited by
the optimized networks persists until $\lambda=1$, when optimization 
neglects the minimization of wiring cost entirely. At this point, the optimized
network uses greater wiring than its WS counterpart, but only slightly.

In contrast to the WS model, a constraint on degree is not maintained
in the optimized model. Figure~\ref{fig:optWD}(b) shows how the
maximum degree increases with $\lambda$ and $p$, for the two models.
The maximum degree, $D$, is normalized by the network size, $n$.
For the optimized network, $D$ is equivalent to the size of the
largest hub. At small $\lambda$, there is no difference between the
two models, but once hubs begin to emerge, $D$ increases sharply for
the optimized network. At both the universal hub, and the near random
graph limit, the maximum degree is $(n-1)$, the size of the universal
hub. In contrast, each edge being rewired only once in the WS model 
allows for only a slight variation in degree.
One also observes a similarity in the variations
of $W$ and $D$ for the optimized networks, since $W$ controls the
size of hubs, and hence $D$.

\subsection{Edge scale distribution}

Since the WS rewiring mechanism exercises no restraint on the length
scales of the rewired edges, the rewired edges are correspondingly
uniformly distributed over the entire length scale range. In
contrast, for the minimally wired networks, lower length scales occur
with a higher probability.

\begin{figure}[!htbp]
\centerline{
 \psfig{figure=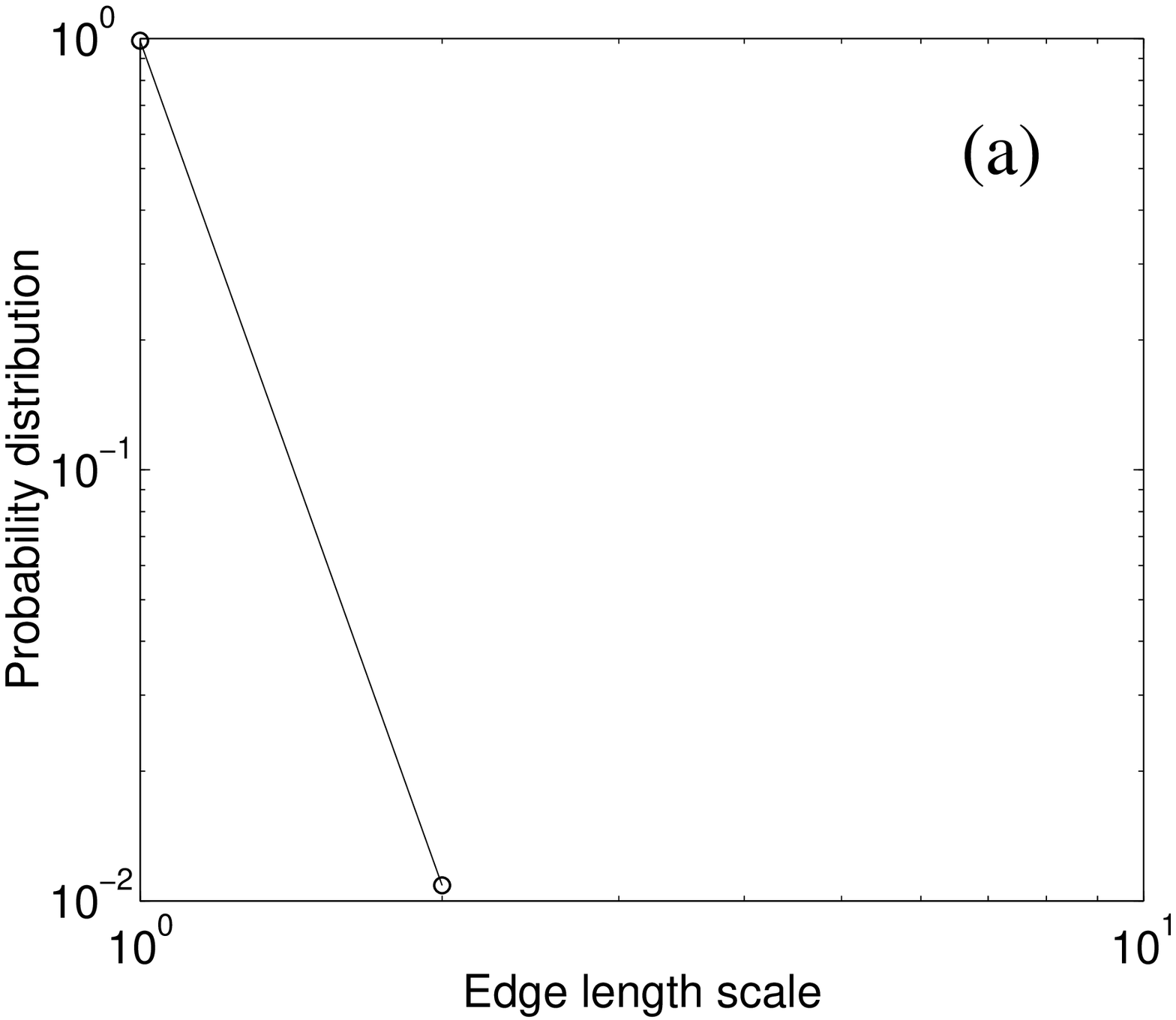,width=6.5cm,height=6.5cm}
 \psfig{figure=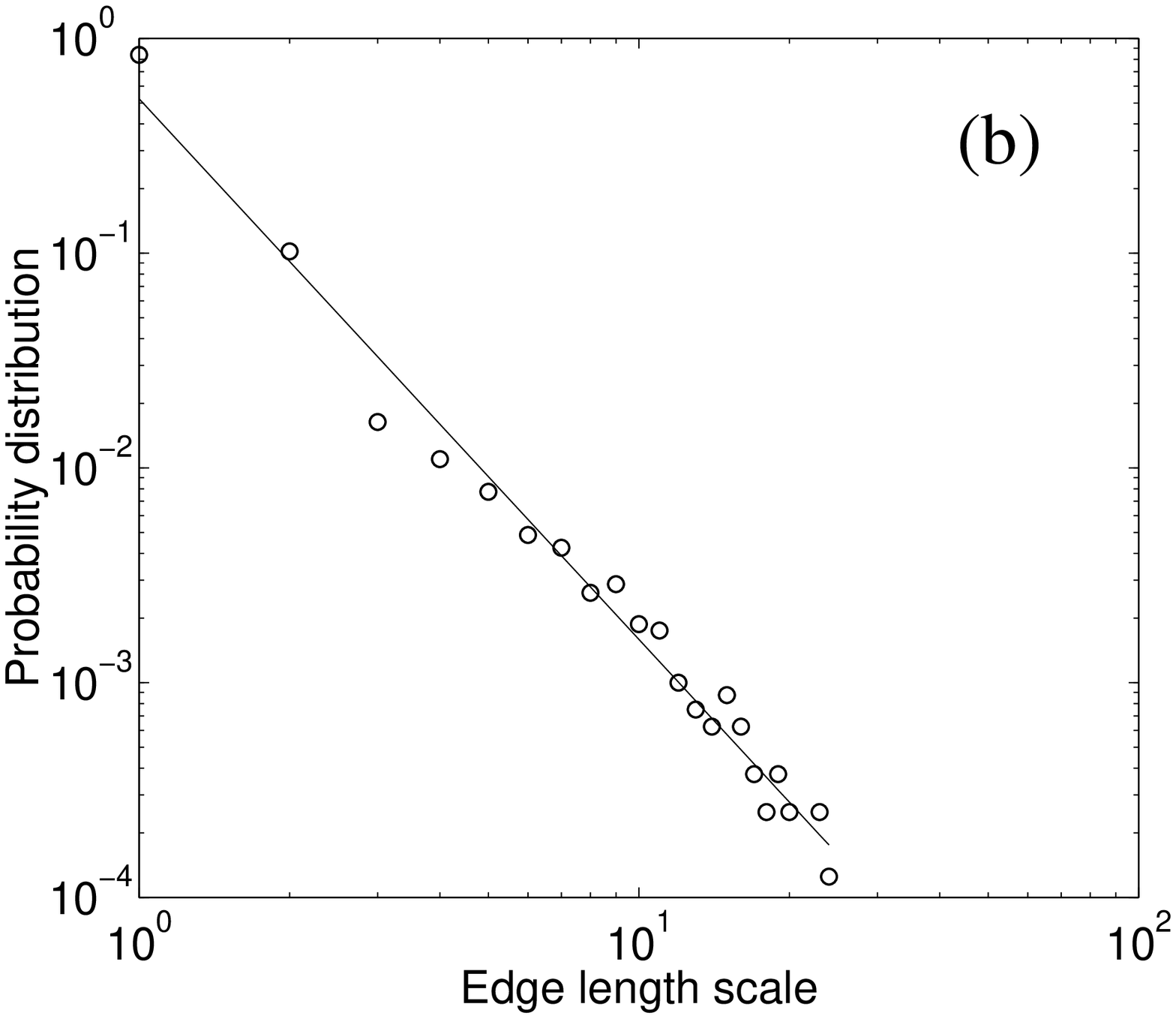,width=6.5cm,height=6.5cm}}
\centerline{
 \psfig{figure=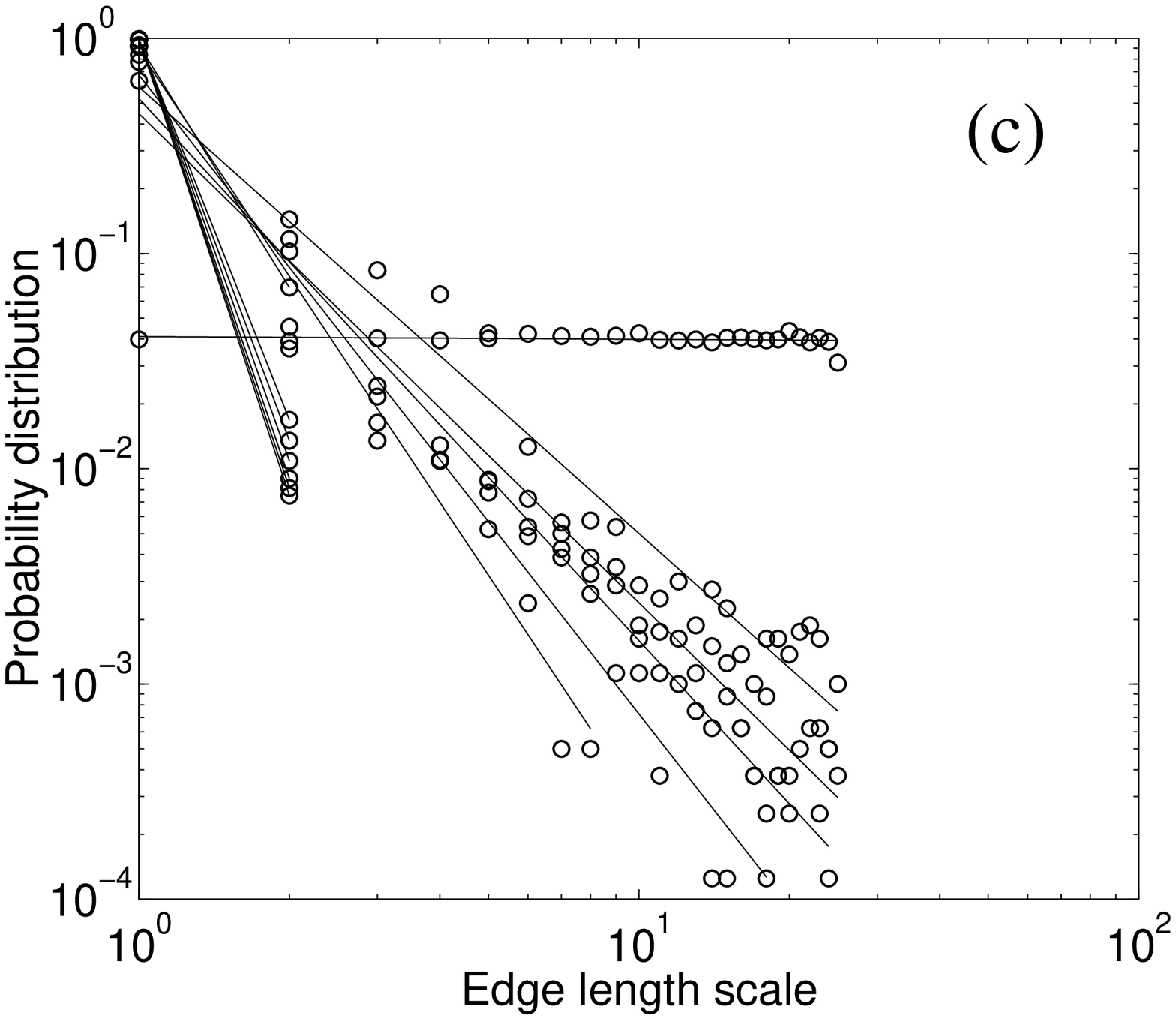,width=6.5cm,height=6.5cm}
 \psfig{figure=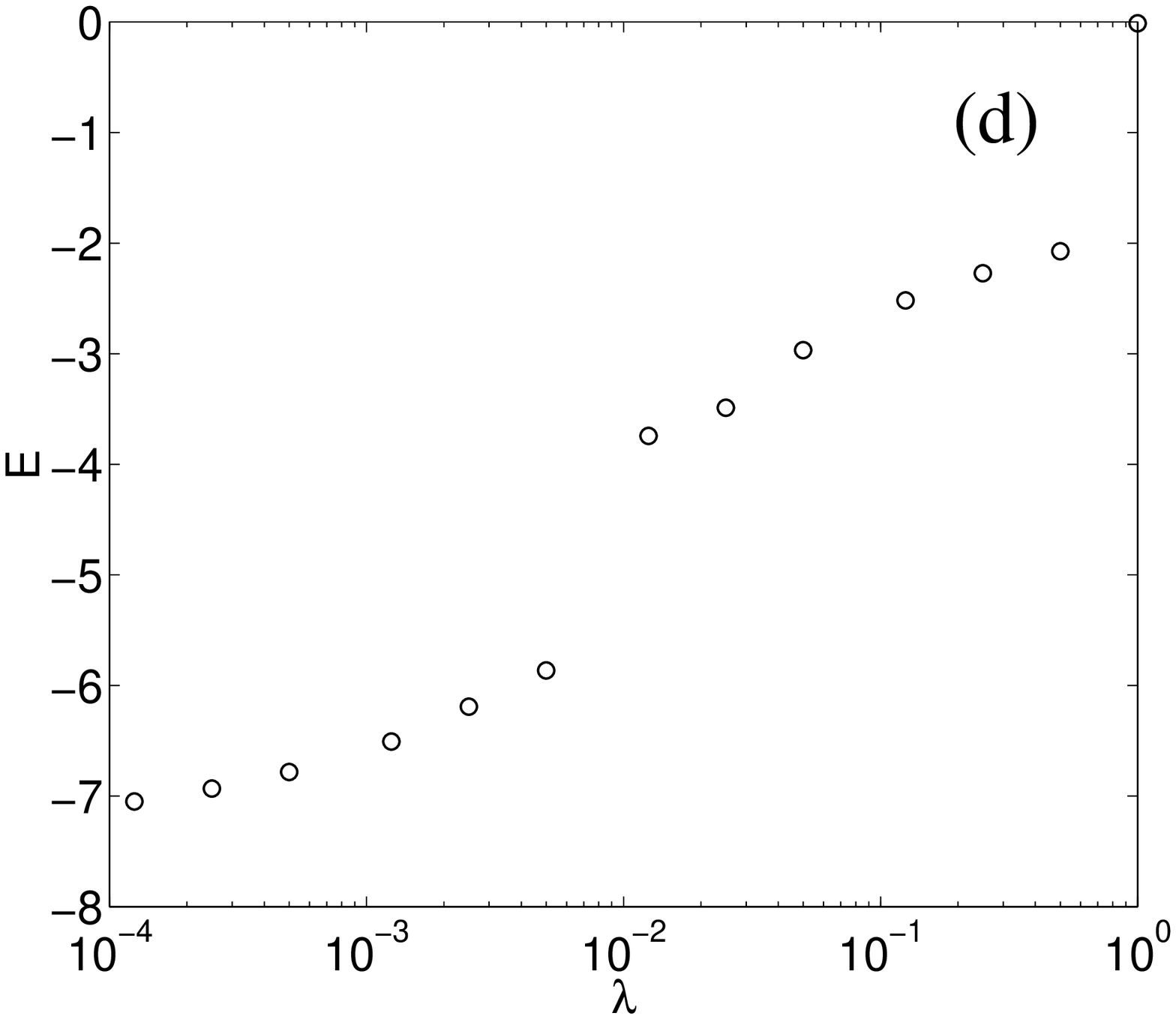,width=6.6cm,height=6.5cm}}
\caption{Log-log plot of the edge scale probability distribution at (a)
$\lambda=0.00125$, and (b) $\lambda=0.125$, for an $n=100, k=4$ optimized 
network. (c) Combined plot of the probability distributions, each with its
associated linear least-squares fit. (d) Variation in the power-law 
exponents with $\lambda$. Each distribution plot is averaged over 40
realizations.}
\label{fig:power-laws}
\end{figure}

Figure~\ref{fig:power-laws} shows plots of the edge scale probability
distribution on a log-log scale, where a power-law behaviour is seen.
Figure~\ref{fig:power-laws}(a) and (b) illustrate the edge scale
distributions at varying $\lambda$, while \ref{fig:power-laws}(c) is
a combined plot which demonstrates the variation in the power-law
distributions with $\lambda$. Each distribution is displayed along
with its associated linear least-squares fit. The variation in their 
exponents, as obtained from the linear least-squares fit to the data 
against $\lambda$, is shown in Fig.\ \ref{fig:power-laws}(d).

The variation in the power law exponents with $\lambda$, can be clearly 
demarcated into two regions. The first, spanning two orders of magnitude 
variation in $\lambda$, exhibits a very slight exponent variation.
Figure~\ref{fig:power-laws}(a) illustrates the typical probability
distribution in this regime. Just two points emerge, since the high
wiring cost constrains almost all edges to have a unit length scale,
with a very slight probability of a higher length scale. A sharp jump
in the exponent marks the beginning of the second regime.
Figure~\ref{fig:power-laws}(b) illustrates the typical probability
distribution in this regime. It is seen that a straight line is a
reasonably good fit to the data over a wide edge scale range. Finally, 
when $\lambda = 1$ and a near random network is achieved, a flat distribution
of length scales results with each scale being equally probable.
The combined plot of all the distributions with their associated
least-squares fits, although noisy, illustrates the behaviour of the 
data (Fig.\ \ref{fig:power-laws}(c)).

The exponent variation clearly reveals two regimes of behaviour. The first jump
in the exponent corresponds to the onset of small-world behaviour, the 
first perceptible reduction in $L$ that is seen in Fig.\ \ref{fig:wsOptL}. 
This marks the beginning of the multiple
scale regime, and is also a signature of hub formation. As we have
mentioned previously, due to computational constraints we were unable
to investigate larger networks. We believe that the noise in the data
is due to the small size of the networks that we have studied.

Finally, we wish to comment upon Kasturirangan's multiple scale
hypothesis. Figure~\ref{fig:power-laws}(d) demonstrates clearly the
connection between the onset of
small-world behaviour and the emergence of multiple length scales in
the network. This clearly supports the claim in \cite{3:kasturi} that
small-worlds arise as a result of the network having connections that
span many length scales, and forms the first 
quantitative support of his hypothesis. It is to
be noted that multiple scales contribute to reducing $L$ in the WS
model as well. However, since no restriction on the length of edges
exists, {\em any} non-zero $p$ will result in multiple scales. Hence,
the onset of small-world behaviour appears with a smooth reduction in $L$.

We have also observed a power law tail for vertex connectivity. We found 
that most vertices had a small degree, and some were well short
of the average degree, with vertices at hub centres gaining at their expense.
Owing to the small network size however, the scaling range was rather limited, 
and so we have not included these results.

\section{Do similar networks exist?}

Any efficient transportation network works under a similar underlying
principle of maximizing connectivity while ensuring that the cost is
minimized. Our results seem to indicate that any efficient
transportation network will be a small-world, and in addition will
exhibit a similar hub connectivity. In a clear illustration of the underlying
principle, any map of airline routes, or roadways shows big cities as being
hubs of connectivity. This is hardly surprising though, since in
such networks, a conscious effort is made toward such a minimization.
However, the same philosophy may well be at work in natural transportation and
other biological networks.

We would also like to point out that our observed hub structure can be seen in
a number of large complex networks ranging from fields as diverse as
the world wide web to the world of actors.
Kleinberg et al. \cite{3:kleinberg} have observed the following
recurrent phenomena on the web:  For any particular topic, there tend
to be a set of ``authoritative'' pages focused on the topic, and a set
of ``hub'' pages, each containing links to useful, relevant pages on
the topic. Also, it has been noted in \cite{3:matthews} that the 
small-world phenomenon in the world of actors arises due to 
``linchpins'': hubs of connectivity 
in the acting industry that transcend genres and eras. In addition,
Barab\'{a}si and Albert \cite{3:barabasi} explore several
large databases describing the topology of large networks that span a
range of fields. They observe that independent of the system and the
identity of its constituents, the probability $P(k)$ that a vertex in
the network interacts with $k$ other vertices decays as a power-law,
following $P(k) \sim k^{-\gamma}$. The power law for the network
vertex connectivity indicates that highly connected vertices (large
$k$) have a large chance of occurring, dominating the connectivity;
hence demonstrating the presence of hubs in these networks. Thus,
hubs seem to constitute an integral structural component of a number
of large and complex random networks, both natural and man-made. 

\section{Conclusions}

Watts and Strogatz showed that small-worlds capture the best of both
graph-worlds: the regular and the random. There has however been no
work citing reasons for their ubiquitous emergence. Our work is an
step is this direction, questioning whether small-worlds can
arise as a tradeoff between optimizing the average degree of
separation between nodes in a network, as well as the total cost of
wiring.

Previous work has concentrated on small-world behaviour that arises as a
result of the random rewiring of a few edges with no constraint of the
length of the edges. On introducing this constraint, we have shown
that an alternate route to small-world behaviour is through the
formation of hubs. The vertex at each hub centre contracts the
distance between every pair of vertices within the hub, yielding a
small characteristic path length. In addition, the introduction of a
hub centre into each neighbourhood serves to sustain the clustering
coefficient at its initially high value. We find that the optimized
networks have $C \ge C_{regular}$, and $L \le L_{random}$ and thus
do better than those described by Watts and Strogatz.

In summary, our work lends support to the idea that a competitive
minimization principle may underly the formation of a small-world
network. Also, we observe that hubs could constitute an integral
structural component of any small-world network, and that power-laws
in edge length scale, and vertex connectivity may be signatures of
this principle in many complex and diverse systems.

Finally, in future work, we will be studying larger networks that were
computationally inaccessible to us at present. We are also
investigating the application of the small-world architecture in the
brain, and also, a dynamic model will be considered to understand the
emergence of small-worlds in social networks.

\begin{center}
{\bf Acknowledgements}
\end{center}
N.M. thanks V. Vinay for very useful discussions.


\begin{thebibliography}{10}
\bibitem{3:watts1} Watts, D. J. and Strogatz, S. H. {\sl Collective
Dynamics of `small world' networks. } Nature, 393:440-442, 1998.
\bibitem{3:barabasi} Barab\'{a}si, A.-L. and Albert, R. {\sl Emergence of
Scaling in Random Networks.} cond-mat/9910332.
\bibitem{3:kasturi} Kasturirangan, R. {\sl Multiple Scales in Small-World
Graphs.} cond-mat/9904055.
\bibitem{3:press} Press, W. H., Teukolsky, S. A., Vetterling, W. T.
and Flannery, B. P. {\sl Numerical Recipes in C: The Art of Scientific
Computing.} Cambridge University Press, Second Edition, 1988.
\bibitem{3:watts2} Watts, D. J. {\sl Small Worlds: The Dynamics of Networks
between Order and Randomness.} Princeton University Press, 1999.
\bibitem{3:barrat} Barrat, A. and Weigt, M. {\sl On the properties of
small-world network models.} cond-mat/9903411.
\bibitem{3:kleinberg} Kleinberg, J. M., Kumar, R., Raghavan, P., 
Rajagopalan, S.
and Tomkins, A. S. {\sl The Web as a Graph: Measurements, Models, and Methods.}
Asano, T. et al. (eds.): COCOON '99, LNCS 1627, 1-17, Springer-Verlag Berlin, Heidelberg, 1999.
\bibitem{3:matthews} Matthews, R. {\sl Get connected.} New Scientist, 4
December 1999.


\end{thebibliography}
\end{document}